\documentclass[showpacs,preprintnumbers,superscriptaddress,amsmath,amssymb,floatfix]{revtex4}
\usepackage[german,english]{babel}
\usepackage{graphicx}
\usepackage{amsmath}
\usepackage{amsfonts}
\usepackage{amssymb}
\usepackage{epsfig}
\usepackage[hang,nooneline]{subfigure}
\newcommand{\insertplot}[5]{\begin{figure}
 \hfill\hbox to 0.05in{\vbox to #5in{\vfill
 \inputplot{#1}{#4}{#5}}\hfill}
 \hfill\vspace{-.1in}
 \caption{#2}\label{#3}
 \end{figure}}
\newcommand{\inputplot}[3]{
 \special{ps: plotfile #1}
}
 
\newcounter{fig}

\begin{document}

\title{\bf Rotating wormhole solutions with a complex phantom scalar field}

\author{Xiao Yan Chew}
\email{xiao.yan.chew@uni-oldenburg.de}
\affiliation{
	Institut f\"ur Physik, Universit\"at Oldenburg, Postfach 2503
	D-26111 Oldenburg, Germany
}

\author{Vladimir Dzhunushaliev}
\email{v.dzhunushaliev@gmail.com}
\affiliation{
	Institute of Experimental and Theoretical Physics,  Al-Farabi Kazakh National University, Almaty 050040, Kazakhstan
}
\affiliation{
	Department of Theoretical and Nuclear Physics,  Al-Farabi Kazakh National University, Almaty 050040, Kazakhstan
}
\affiliation{
	Academician J. Jeenbaev Institute of Physics of the NAS of the Kyrgyz Republic, 265 a, Chui Street, Bishkek 720071,  Kyrgyz Republic
}
\affiliation{
	Institut f\"ur Physik, Universit\"at Oldenburg, Postfach 2503
	D-26111 Oldenburg, Germany
}

\author{Vladimir Folomeev}
\email{vfolomeev@mail.ru}
\affiliation{
	Institute of Experimental and Theoretical Physics,  Al-Farabi Kazakh National University, Almaty 050040, Kazakhstan
}
\affiliation{
	Academician J. Jeenbaev Institute of Physics of the NAS of the Kyrgyz Republic, 265 a, Chui Street, Bishkek 720071,  Kyrgyz Republic
}
\affiliation{
	Institut f\"ur Physik, Universit\"at Oldenburg, Postfach 2503
	D-26111 Oldenburg, Germany
}

\author{Burkhard Kleihaus}
\email{b.kleihaus@uni-oldenburg.de}
\affiliation{
	Institut f\"ur Physik, Universit\"at Oldenburg, Postfach 2503
	D-26111 Oldenburg, Germany
}

\author{Jutta Kunz}
\email{jutta.kunz@uni-oldenburg.de}
\affiliation{
	Institut f\"ur Physik, Universit\"at Oldenburg, Postfach 2503
	D-26111 Oldenburg, Germany
}

\date{\today}

\begin{abstract}
We consider rotating wormhole solutions supported by a
complex phantom scalar field with a quartic self-interaction,
where the phantom field induces the rotation of the spacetime.
The solutions are regular and asymptotically flat.
A subset of solutions describing static but not spherically symmetric 
wormholes is also obtained.
\end{abstract}

\maketitle

\section{Introduction}

Wormholes have received much attention in recent years. 
Numerous investigations have been performed
addressing the possible signatures of wormholes
in astrophysical searches,
including gravitational lensing by wormholes
\cite{Cramer:1994qj,Safonova:2001vz,Perlick:2003vg,Nandi:2006ds,Abe:2010ap,Toki:2011zu,Nakajima:2012pu,Tsukamoto:2012xs,Kuhfittig:2013hva,Bambi:2013nla,Takahashi:2013jqa,Tsukamoto:2016zdu},
shadows of wormholes
\cite{Bambi:2013nla,Nedkova:2013msa,Ohgami:2015nra,Shaikh:2018kfv,Gyulchev:2018fmd},
or accretion disks around wormholes
\cite{Harko:2008vy,Harko:2009xf,Bambi:2013jda,Zhou:2016koy,Lamy:2018zvj}.
Particular emphasis has been placed on the question as
to what extent wormholes might mimick black holes
\cite{Damour:2007ap,Bambi:2013nla,Azreg-Ainou:2014dwa,Dzhunushaliev:2016ylj,Cardoso:2016rao,
      Konoplya:2016hmd,Nandi:2016uzg,Bueno:2017hyj,Blazquez-Salcedo:2018ipc}.

As discussed by Morris and Thorne \cite{Morris:1988cz,Morris:1988tu}
the presence of exotic matter allows for the construction of wormholes
in General Relativity.
The simplest such wormhole solutions are the Ellis wormholes,
which are obtained with a real massless phantom field
\cite{Ellis:1973yv,Bronnikov:1973fh,Kodama:1978dw,Ellis:1979bh,Lobo:2005us,Lobo:2017oab},
i.e., a scalar field whose kinetic term has the opposite sign
as compared to ordinary scalar fields.
The static Ellis wormholes are known in closed form.
Their rotating generalizations, however, 
are either known perturbatively \cite{Kashargin:2007mm,Kashargin:2008pk}
or numerically
\cite{Kleihaus:2014dla,Chew:2016epf,Kleihaus:2017kai}.

Whereas the static Ellis wormholes can be chosen to be symmetric
with respect to reflection at the throat such that both parts of
the spacetime are completely alike, the presence of rotation
necessarily breaks this symmetry for Ellis wormholes
\cite{Kashargin:2007mm,Kashargin:2008pk,Kleihaus:2014dla,Chew:2016epf,Kleihaus:2017kai}.
(For the discussion of rotating Ellis wormholes in Scalar-Tensor Theories see 
\cite{Chew:2018vjp}).
The presence of further ordinary fields, however, can allow
for reflection symmetric rotating wormholes,
as recently shown for the case of an ordinary complex scalar field
\cite{Hoffmann:2017vkf}.
In fact, in these configurations the rotation of the wormhole is
induced by the rotation of the matter fields.

Besides a real phantom field one can, however, also consider
a complex phantom field. In that case one can try to impose
rotation directly on the complex phantom field, 
and thus obtain rotating wormhole configurations that
are symmetric and do not need any additional matter fields.
This is the goal of the present work.

Non-rotating wormholes based on a complex phantom field  with a {\sl Mexican hat} type potential
have been considered before \cite{Dzhunushaliev:2017syc}.
The $U(1)$ symmetry of the theory leads to a conserved
current associated with a conserved charge, the particle number.
As in boson stars
\cite{Jetzer:1991jr,Lee:1991ax,Schunck:2003kk,Liebling:2012fv}
the phantom field of the wormholes possesses a harmonic time-dependence,
while their metric is static and spherically symmetric.
However, there exist also wormhole solutions, where the time-dependence
of the phantom field vanishes together with the particle number.
For these solutions the complex phantom field reduces to a real valued field.

Here we impose rotation on the complex phantom field
in the same way that rotation is imposed for an ordinary complex
scalar field in the construction of boson stars 
\cite{Schunck:1996,Schunck:1996he,Ryan:1996nk,Yoshida:1997qf,Schunck:1999pm,Kleihaus:2005me,Kleihaus:2007vk}
or of wormholes immersed in rotating bosonic matter
\cite{Hoffmann:2017vkf,Hoffmann:2018oml}.
Thus the ansatz for the complex phantom field has an explicit
dependence on the azimuthal angle featuring 
an integer $n$ for spatial single-valuedness of the solutions,
as in the well-known case of the spherical harmonics.
The rotating solutions then do not only possess a mass
and a particle number,
but they also carry an angular momentum proportional
to the particle number with proportionality constant $n$,
a relation well-known from boson stars.

The paper is organized as follows.
In section II we present the action, Ans\"atze, field equations
and boundary conditions, and we define various physical quantities.
In section III we present the solutions. 
We first briefly recall the non-rotating case, and then discuss
the properties of the rotating solutions,
where we analyze the global charges and the wormhole geometries.
We give our conclusions in section IV.

\section{Theoretical setting}

In this section we present the action,
the Ans\"atze for the metric and the phantom field, 
the resulting field equations and the boundary conditions. 
Subsequently we define the global charges and
the geometrical properties of the wormholes.

\subsection{Action}

The action consists of the Einstein-Hilbert action
and the action for a complex phantom field
\begin{equation}
S=\int \left[ \frac{1}{2 \kappa}{\cal R} 
+{\cal  L}_{\rm ph} \right] \sqrt{-g}\  d^4x  \ .
 \label{action}
\end{equation}
Here $\cal R$ is the curvature scalar, $\kappa=8\pi G$ is the
coupling constant, $g$ denotes the determinant of the metric,
and ${\cal L}_{\rm ph}$ represents
the Lagangian of the complex phantom field $\Phi$
\begin{equation}
{\cal L}_{\rm ph} = 
\frac{1}{2} g^{\mu\nu}\left( \partial_\mu\Phi^* \partial_\nu\Phi
                            + \partial_\nu\Phi^* \partial_\mu\Phi 
 \right) - V(|\Phi|^2)\ ,
\label{lphi}
\end{equation}
where the asterisk represents complex conjugation
and  the potential $V$
\begin{equation}
V(|\Phi|^2) = -m_{\rm ph}^2 |\Phi|^2 +\Lambda  |\Phi|^4
\label{pot}
\end{equation}
consists of a mass term with boson mass parameter $m_{\rm ph}$ 
and  a quartic self-interaction term with  coupling parameter
$\Lambda$. For the discussion of the potential see \cite{Dzhunushaliev:2017syc}.

Variation of the action with respect to the metric and the phantom field
lead to the Einstein equations
\begin{equation}
G_{\mu\nu}= {\cal R}_{\mu\nu}-\frac{1}{2}g_{\mu\nu}{\cal R} =  \kappa T_{\mu\nu}
\label{ee} 
\end{equation}
with stress-energy tensor
\begin{equation}
T_{\mu\nu} = g_{\mu\nu}{{\cal L}}_{\rm ph}
-2 \frac{\partial {{\cal L}}_{\rm ph}}{\partial g^{\mu\nu}} \ ,
\label{tmunu} 
\end{equation}
and to the phantom field equation
\begin{equation}
\nabla^\mu \nabla_\mu \Phi 
   = 
   m_{\rm ph}^2 \Phi -2\Lambda  |\Phi|^2 \Phi .
\label{ephi} 
\end{equation}

It is convenient to introduce dimensionless quantities
\begin{equation}
\hat{\Phi} = \sqrt{\kappa}  \Phi \ , \ \ \ 
\hat{\eta} = \frac{1}{\lambda} \eta \ ,
\label{scaling}
\end{equation}
%
where $\eta$ is the radial coordinate (see Eq.~(\ref{lineel})).
Consequently,
\begin{equation}
\hat{G}_{\mu\nu} =  \hat{T}_{\mu\nu} \ ,
\label{ees} 
\end{equation}
where the potential
$-\hat{m}_{\rm ph}^2 |\hat{\Phi}|^2+\hat{\Lambda}|\hat{\Phi}|^4$ on the right hand side
contains the dimensionless parameters  
\begin{equation}
\hat{m}_{\rm ph}=\left(\frac{m_{\rm ph}}{m_{\rm P}}\right)\left(\frac{M_0}{m_{\rm P}}\right) \ , \ \ \ 
\hat{\Lambda} = \left(\frac{M_0}{m_{\rm P}}\right)^2\frac{1}{8\pi} \Lambda \ .
\label{hatpara} 
\end{equation}
Here $m_{\rm P}$ denotes the Planck mass,
and the mass scale $M_0$ is related 
to the length scale by $M_0 = \lambda /G$.
We will omit the hats
in the following for reasons of notational simplicity.

\subsection{Ans\"atze}

To incorporate the non-trivial topology we employ as line element for the
metric
\begin{equation}
ds^2 = -e^{f} dt^2 
    +e^{q-f}\left[e^b(d\eta^2 + h d\theta^2)+ h \sin^2\theta
    (d\varphi -\omega dt)^2\right] \ ,
\label{lineel}
\end{equation}
where the functions $f$, $q$, $b$ and $\omega$ depend on
the radial coordinate $\eta$ and the polar angle $\theta$,
and the  auxiliary function $h = \eta^2 +\eta_0^2$ contains the
throat parameter $\eta_0$.
The radial coordinate $\eta$ covers the interval
$-\infty< \eta < \infty$, 
where both limits $\eta\to \pm\infty$
represent asymptotically flat regions.

For the complex phantom field
we adopt the Ansatz
\begin{equation}
\Phi(t,\eta,\theta, \varphi) 
  =  \phi (\eta,\theta) ~ e^{ i\omega_s t +  i n \varphi} \    \label{ansatzp}
\end{equation}
with the real function $\phi (\eta,\theta)$, 
the real boson frequency $\omega_s$
and the integer winding number $n$.
This Ansatz agrees with the one employed 
for rotating $Q$-balls and boson stars
\cite{Schunck:1996,Schunck:1996he,Ryan:1996nk,Yoshida:1997qf,Schunck:1999pm,Kleihaus:2005me,Kleihaus:2007vk}
and for wormholes immersed in rotating bosonic matter
\cite{Hoffmann:2017vkf,Hoffmann:2018oml}.
Non-rotating, spherically symmetric solutions can be obtained with the Ansatz 
Eqs.~(\ref{lineel}) and (\ref{ansatzp}) for $n=0$ with $b=0$, $\omega=0$,
and the remaining functions depending on $\eta$ only.

\subsection{Einstein and Matter Field Equations}

By substituting the Ans\"atze (\ref{lineel}) and (\ref{ansatzp})
into the set of Einstein equations 
$E_\mu^\nu=G_\mu^\nu- T_\mu^\nu=0$, 
we find the following set of equations
\begin{equation}
f_{,\eta\eta} 
+ \frac{f_{,\theta\theta}}{h} 
+ f_{,\eta} \frac{h q_{,\eta} + 4 \eta}{2 h} 
+f_{,\theta} \frac{2 \cot\theta + q_{,\theta} }{2 h} 
- e^{q-2f} \sin^2\theta \left(h \omega_{,\eta}^2 + \omega_{,\theta}^2\right)
 =  
-2 \left[2 (n \omega + \omega_s)^2\phi^2 + V(\phi^2)  e^{f} \right] e^{b+q-2f}  \ ,
\label{eq1}
\end{equation}
\begin{equation}
q_{,\eta\eta} 
+ \frac{q_{,\theta\theta}}{h}
+\frac{q_{,\eta}^2}{2} 
+ \frac{3\eta q_{,\eta}}{h} 
+ \frac{q_{,\theta}^2}{2 h} 
+ \frac{2 q_{,\theta} \cot\theta}{h} 
 = 
-4 \left(
  e^{q-2f} h\left[(n \omega + \omega_s)^2\phi^2 +  V(\phi^2) e^{f} \right]\sin^2\theta 
- n^2 \phi^2
\right)
\frac{e^{b}}{h \sin^2\theta} \ ,
\label{eq2}
\end{equation}
\begin{eqnarray}
& &
b_{,\eta\eta} 
+ \frac{b_{,\theta\theta}}{h} 
+b_{,\eta} \frac{\eta}{h} 
\nonumber\\
&  &
+ \frac{1}{2}\left[
 f_{,\eta}^2+\frac{f_{,\theta}^2}{h}-q_{,\eta}^2-\frac{q_{,\theta}^2}{h}
 - \frac{4}{h} \left( \cot\theta  q_{,\theta} + \eta q_{,\eta} \right) 
+ 
 \frac{4}{h}\left( 1-\frac{\eta^2}{h}\right)
- 3 e^{q-2f} h \left(\omega_{,\eta}^2 +\frac{\omega_{,\theta}^2}{h}\right)\sin^2\theta 
\right]
\nonumber\\
& = &
2 \left(\phi_{,\eta}^2 + \frac{\phi_{,\theta}^2}{h} \right)
+2 
\left( e^{q-2f} h \left[(n \omega + \omega_s)^2\phi^2 + V(\phi^2) e^f\right]\sin^2\theta 
- 3n^2\phi^2
\right)
\frac{e^{b}}{ h \sin^2\theta} \ ,
\label{eq3}
\end{eqnarray}
\begin{equation}
\omega_{,\eta\eta} 
+ \frac{\omega_{,\theta\theta}}{h}
+\frac{\omega_{,\eta}}{2} \left( 3q_{,\eta}- 4 f_{,\eta} + \frac{8\eta}{h} \right)
+\frac{\omega_{,\theta} }{2h}
\left(3 q_{,\theta} - 4 f_{,\theta} +6 \cot\theta \right)
  =  
 - 4 n (n\omega + \omega_s)\phi^2\frac{e^b}{h\sin^2\theta} \ ,
\label{eq4}
\end{equation}
resulting from $E^t_t=0$, $E^\eta_\eta+E^\theta_\theta=0$, 
$E^\varphi_\varphi=0$ and  $E^t_\varphi=0$, respectively.
From $d:=E^\eta_\eta-E^\theta_\theta$ we find in addition the condition
\begin{eqnarray}
d  = 0 
  & = &
 q_{,\eta\eta} - \frac{1}{h}q_{,\theta\theta} 
- \left( \omega_{,\eta}^2- \frac{1}{h}\omega_{,\theta}^2 \right) h \sin^2\theta e^{-2f+q}
- \frac{1}{2}\left(q_{,\eta}^2 - \frac{1}{h}q_{,\theta}^2\right) 
+\left(f_{,\eta}^2- \frac{1}{h}f_{,\theta}^2\right)
\nonumber\\ & &
-  \frac{1}{h}\left(q_{,\eta} \left(h b_{,\eta} + \eta\right) -q_{,\theta} b_{,\theta}\right)
-\frac{2}{h^2 \sin\theta}\left[ 
 \left(\eta \sin\theta b_{,\eta} - \cos\theta b_{,\theta} \right) h
- 2 \eta_0^2 \sin\theta
\right]
\nonumber\\ & &
-4 \left( \phi_{,\eta}^2 - \frac{1}{h}\phi_{,\theta}^2\right) \ .
\label{constrd}
\end{eqnarray}
to which we refer to as constraint.
The field equation for the phantom field function $\phi(\eta,\theta)$
is obtained from Eq.~(\ref{ephi}),
%
%
%
\begin{equation}
  \phi_{,\eta\eta} 
+ \frac{\phi_{,\theta\theta}}{h}
+\frac{\phi_{,\eta}}{2} \left( q_{,\eta} + 4 \frac{\eta}{h}\right)
+ \frac{\phi_{,\theta} }{2h} \left( q_{,\theta} +2 \cot\theta \right)
+\left( 
e^{q-2f}h\left[(n \omega+ \omega_s )^2
+ \frac{d V(\phi^2)}{d \phi^2}  e^{f} \right]\sin^2\theta
- n^2
\right)\phi\frac{e^{b} }{h \sin^2\theta} = 0 \ .
\label{eq5}
\end{equation}

Inspection of the system of equations shows 
that it is symmetric with respect to reflection of the
radial coordinate, $\eta \to - \eta$.
Consequently reflection symmetric solutions will exist,
although non-symmetric solutions might exist as well.
Since in this study we will consider only reflection symmetric solutions,
it is sufficient to restrict the computations to 
the interval $0\leq \eta < \infty$, 
where $\eta=0$ corresponds to the wormhole throat.

We note that $\eta_0$ is not a free parameter,
when $\omega_s$ and $\Lambda$ are freely varied.
Instead, for any value of $\omega_s$ and $\Lambda$,
$\eta_0$ has to be chosen such 
that the constraint Eq.~(\ref{constrd}) is satisfied.

\subsection{Boundary Conditions}

Let us now specify the boundary conditions employed
to solve the above set of five coupled partial differential equations (PDEs)
of second order. 
In particular, we need to impose conditions for each function
at the boundaries of the domain of integration.
These consist of the throat at $\eta=0$, 
the axis of rotation $\theta = 0$,
the equatorial plane $\theta = \pi/2$,
and the asymptotic region $\eta \to \infty$.

Guided by symmetry arguments, we
now specify our detailed choice of boundary conditions.
Asking for reflection symmetry means that
the normal derivatives of all functions have to
vanish at the throat, 
\begin{equation}
\left. \partial_\eta f(\eta,\theta)\right|_{\eta =0} = 
\left. \partial_\eta q(\eta,\theta)\right|_{\eta =0} = 
\left. \partial_\eta b(\eta,\theta)\right|_{\eta =0} = 
\left. \partial_\eta \omega(\eta,\theta)\right|_{\eta=0 } = 
\left. \partial_\eta \phi(\eta,\theta)\right|_{\eta=0 } = 0 \ .
\label{bc_throat}
\end{equation}
We demand that the metric should be asymptotically flat
and the phantom field should vanish asymptotically
\begin{equation}
\left. f(\eta,\theta)\right|_{\eta \to \infty} = 
\left. q(\eta,\theta)\right|_{\eta \to \infty} = 
\left. b(\eta,\theta)\right|_{\eta \to \infty} = 
\left. \omega(\eta,\theta)\right|_{\eta \to \infty} = 
\left. \phi(\eta,\theta)\right|_{\eta \to \infty} = 0 \ .
\label{bc_pminfty}
\end{equation}
Reflection symmetry with respect to the equatorial plane leads to
the conditions
\begin{equation}
\left.\partial_\theta f(\eta,\theta)\right|_{\theta = \frac{\pi}{2}} = 
\left.\partial_\theta q(\eta,\theta)\right|_{\theta = \frac{\pi}{2}} = 
\left.\partial_\theta b(\eta,\theta)\right|_{\theta = \frac{\pi}{2}} = 
\left.\partial_\theta \omega(\eta,\theta)\right|_{\theta = \frac{\pi}{2}} =
\left.\partial_\theta \phi(\eta,\theta)\right|_{\theta = \frac{\pi}{2}} = 0 \ .
\label{bc_eqplane}
\end{equation}
Regularity along the rotation axis requires
\begin{equation}
\left.\partial_\theta f(\eta,\theta)\right|_{\theta = 0} = 
\left.\partial_\theta q(\eta,\theta)\right|_{\theta = 0} = 
\left.\partial_\theta \omega(\eta,\theta)\right|_{\theta = 0} = 0 \ , \ \ \
\left. b(\eta,\theta)\right|_{\theta = 0} = 0 \ , \ \ \
\left. \phi(\eta,\theta)\right|_{\theta = 0} =0  \ .
\label{bc_axis}
\end{equation}
For the non-rotating solutions the last boundary condition has to be replaced
by $\left.\partial_\theta  \phi(\eta,\theta)\right|_{\theta = 0} =0$.

\subsection{Mass, angular momentum and particle number}

Let us now address the global charges of the solutions.
The (dimensionless) mass $M$ and the (dimensionless) angular momentum $J$ can be obtained from 
the corresponding Komar integrals and read off
from the asymptotic behaviour of the metric functions
\begin{equation}
f \longrightarrow -\frac{2 M}{\eta} \ , \ \ \ 
\omega \longrightarrow \frac{2 J}{\eta^3}  \ \ \ \ {\rm as} \ \eta \to \infty \ .
\label{MJinfty}  
\end{equation}
The Komar integrals can also be transformed with the help of the
Einstein equations to yield
\begin{equation}
\label{intm}
M=\int(2T^t_t-T^\mu_\mu)\sqrt{-g}drd\theta d\varphi \ , 
\end{equation}
and
\begin{equation}
\label{intj}
J=-\int T^t_\varphi\sqrt{-g}drd\theta d\varphi \ .
\end{equation}

The particle number $Q$ is a Noether charge, associated with 
the conserved $U(1)$ current,
\begin{equation}
\label{intq}
Q = \int_\Sigma  j_\mu n^\mu dV \ ,  
\end{equation}
where the integrand corresponds to $-j^t\sqrt{-g}$.
As noted by Schunck and Mielke \cite{Schunck:1996,Schunck:1996he} in the case of boson stars, 
$T^t_\varphi=n j^t$.
By comparing Eqs.~(\ref{intq}) and (\ref{intj}) 
one then finds the relation
between angular momentum and particle number
\cite{Schunck:1996,Schunck:1996he,Ryan:1996nk,Yoshida:1997qf,Schunck:1999pm,Kleihaus:2005me,Kleihaus:2007vk}
\begin{equation}
\label{QJrelation}
J=nQ \ .
\end{equation}
This relation is also known to hold for symmetric wormholes
immersed in rotating bosonic matter \cite{Hoffmann:2017vkf},
and it also holds for the symmetric wormholes with a complex phantom field
studied here.

\subsection{Wormhole throat}

The most important geometrical property of the solutions is their throat.
To analyze the throat structure we consider the
circumferential radius $R_e(\eta)$ in the equatorial plane,
\begin{equation}
R_e(\eta) = \sqrt{h}\left. e^{\frac{q-f}{2}}\right|_{\theta = \pi/2} \ .
\label{Rade}  
\end{equation}
Clearly for large $\eta$ the circumferential radius $R_e(\eta)$
diverges. However, for $\eta \to 0$ the circumferential radius
reaches a minimum. The minimal surface at $\eta=0$ therefore
corresponds to the throat of the wormhole solution.

In principle, more than one minimum could exist, and in between
the minima there might arise local maxima. In that case
the spacetime would have multiple throats with
equators in between.
At $\eta=0$ there will always either be a throat or an equator,
when the solutions possess reflection symmetry.
%

\subsection{Ergosurface and static orbits}

The wormhole spacetimes could also feature an ergosurface,
\begin{equation}
g_{tt}(\eta,\theta) = 0 \ .
\label{ergobound}  
\end{equation}
The condition $g_{tt}>0$ then defines the ergoregion, whose boundary
is the ergosurface.

In \cite{Collodel:2017end} is was shown that in stationary rotating spacetimes static orbits 
in the equatorial plane may exist. On this kind of orbit a particle stays at rest relative to an observer in
the asymptotic region when it was at rest initially. The necessary and sufficient
condition for a static orbit is that $g_{tt}$ possesses a local maximum (stable) or
a local minimum (unstable) in some region where $g_{tt}<0$. It was demonstrated in \cite{Collodel:2017end} that static orbits 
exist for rotating boson stars and wormholes immersed in rotating matter. 
For symmetric wormhole solutions considered in this work $g_{tt}$ always is extremal at the throat. 
Hence a static orbit always resides at the throat.

\section{Results}

Let us now analyze the properties of the wormhole solutions 
based on a complex phantom field.
We will first recall the non-rotating case,
and subsequently we will discuss the rotating case.
The wormhole solutions depend on three continuous parameters, 
represented by 
the boson mass parameter  $m_{\rm ph}$,
the boson frequency $\omega_s$,
and the quartic self-interaction strength $\Lambda$,
and in addition on the integer winding number $n$,
which must be non-zero in the case of rotation.

We note that the set of coupled Einstein and phantom field equations
are invariant under a scaling transformation, i.e.,
\begin{equation}
\eta \to \lambda \eta \ , \ \ \ 
\eta_0 \to \lambda \eta_0 \ , \ \ \ 
\omega \to \frac{1}{\lambda} \omega \ , \ \ \ 
\omega_s \to \frac{1}{\lambda} \omega_s \ , \ \ \ 
m_{\rm ph} \to \frac{1}{\lambda} m_{\rm ph} \  , \ \ \ 
\Lambda \to \frac{1}{\lambda^2} \Lambda \ .
\end{equation}
To break this scaling invariance
we choose for the boson mass parameter the value $m_{\rm ph}=1$.
The remaining free parameters are then 
the boson frequency $\omega_s$, 
the coupling constant $\Lambda$, and the winding number $n$.

\subsection{Non-rotating solutions}

Let us start with the non-rotating wormhole solutions,
which are obtained for vanishing winding number $n=0$.
The Ansatz and the field equations then simplify considerably,
and a system of non-linear coupled ODEs is obtained.
The properties of the non-rotating wormhole solutions 
have been investigated in \cite{Dzhunushaliev:2017syc}.
Here we will summerize them briefly in order
to be able to compare with the rotating case.

\begin{figure}[p!]
\begin{center}
\mbox{\hspace{0.2cm}
\subfigure[][]{\hspace{-1.0cm}
\includegraphics[height=.20\textheight, angle =0]{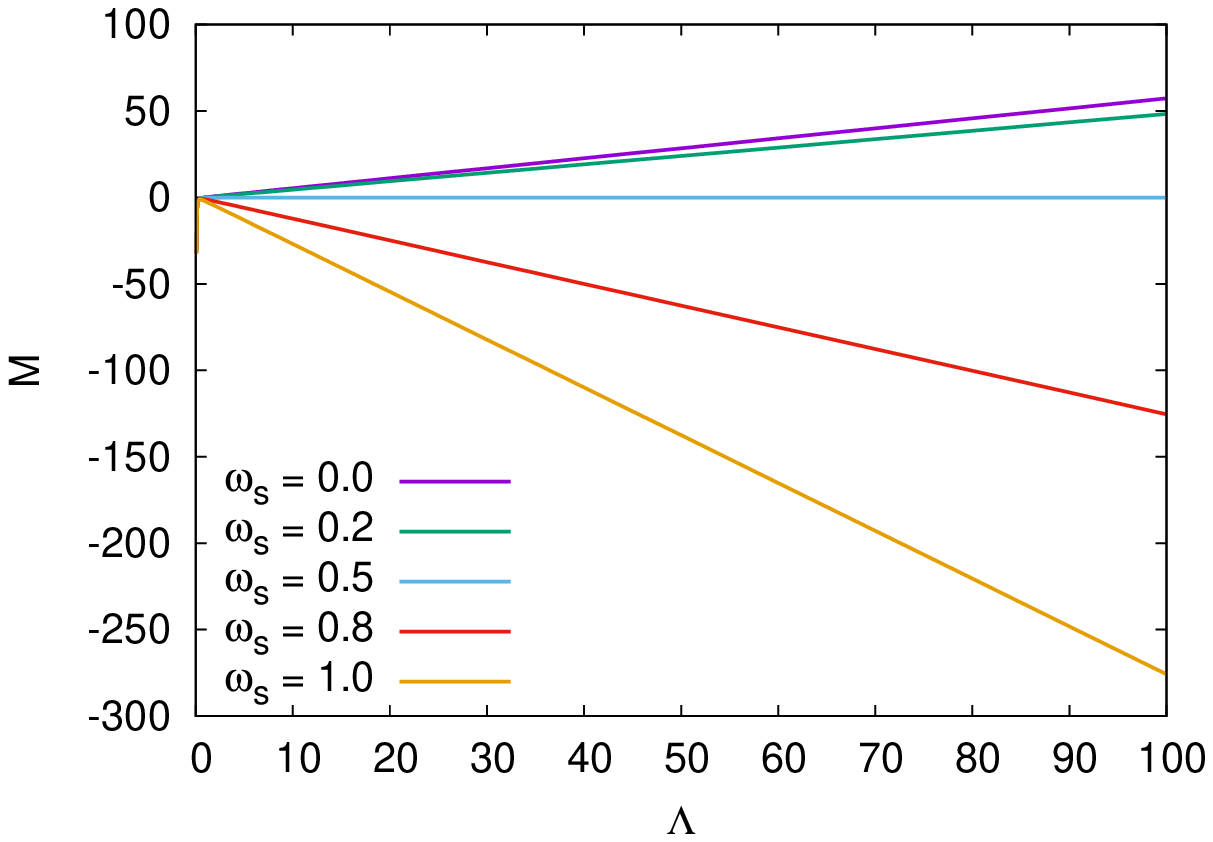}
\label{Fig1a}
}
\subfigure[][]{\hspace{-0.5cm}
\includegraphics[height=.20\textheight, angle =0]{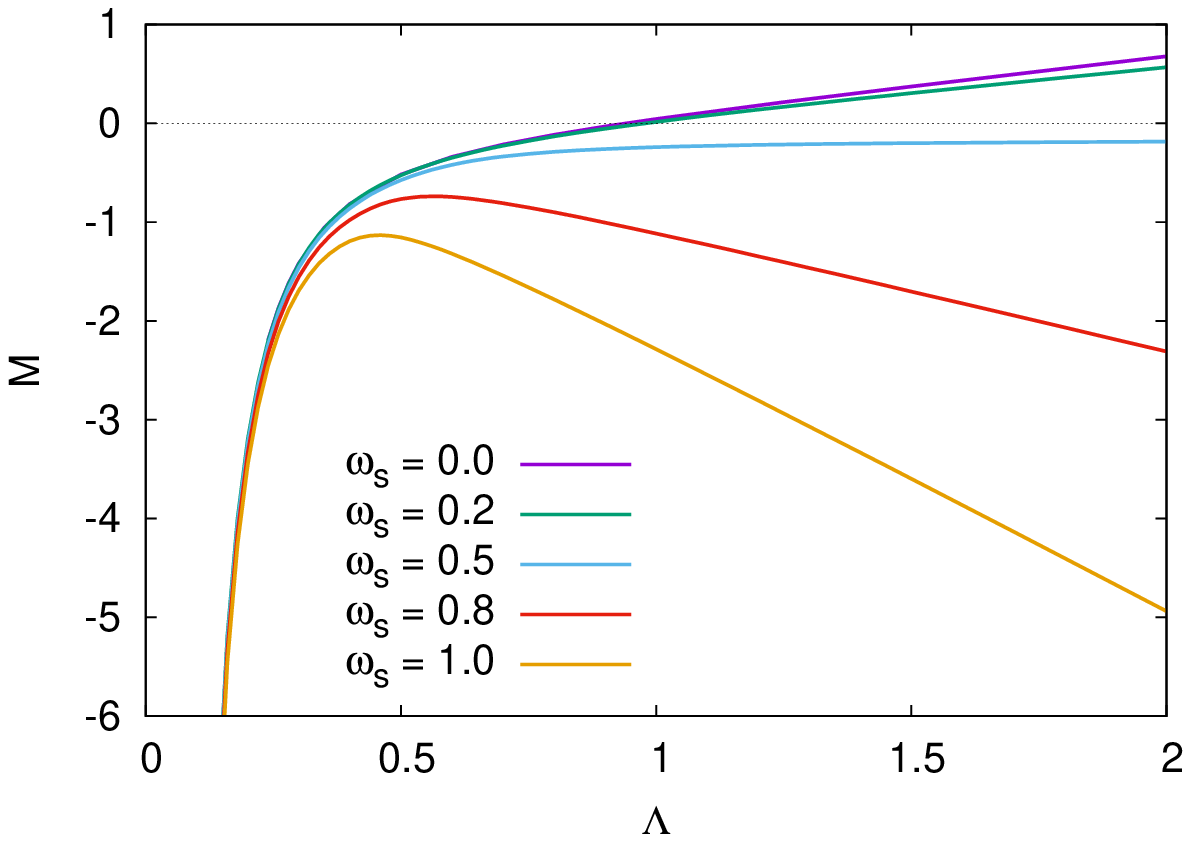}
\label{Fig1b}
}
}
\mbox{\hspace{0.2cm}
\subfigure[][]{\hspace{-1.0cm}
\includegraphics[height=.20\textheight, angle =0]{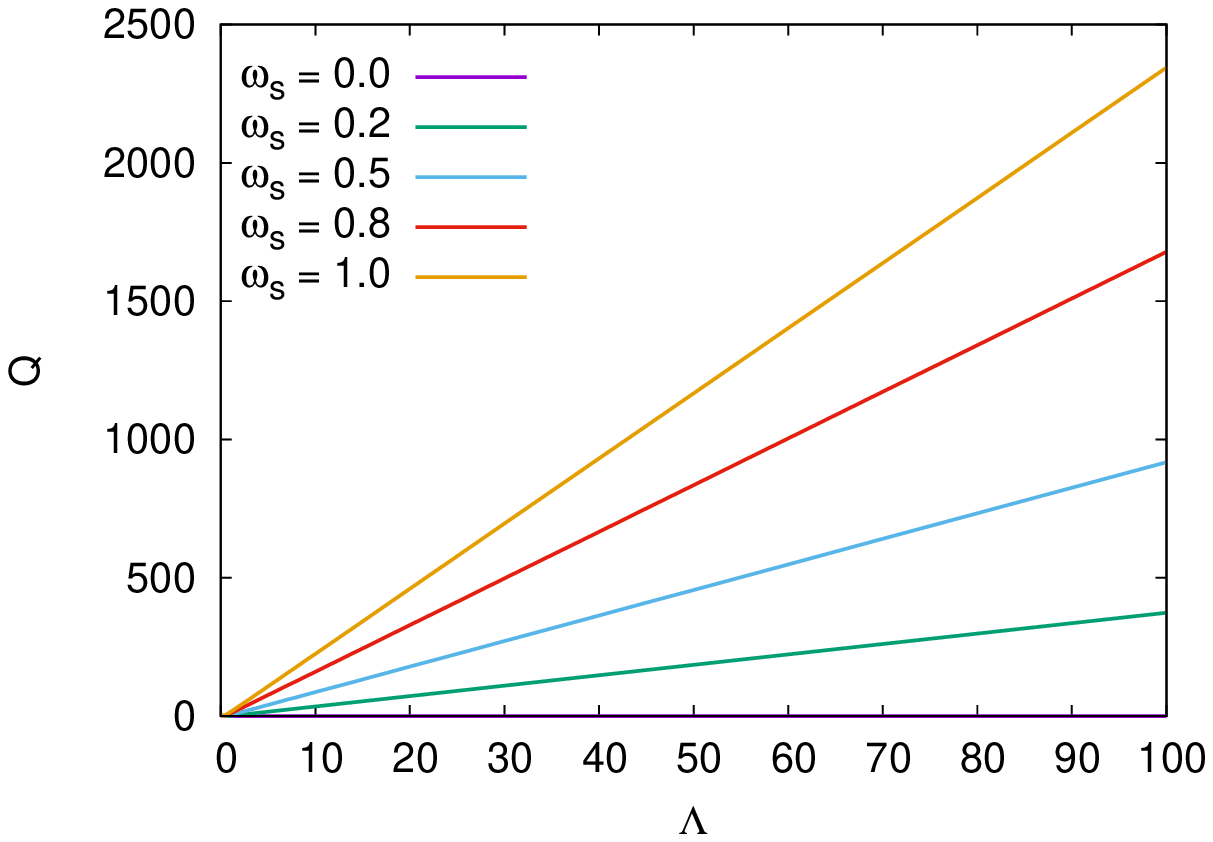}
\label{Fig1c}
}
\subfigure[][]{\hspace{-0.5cm}
\includegraphics[height=.20\textheight, angle =0]{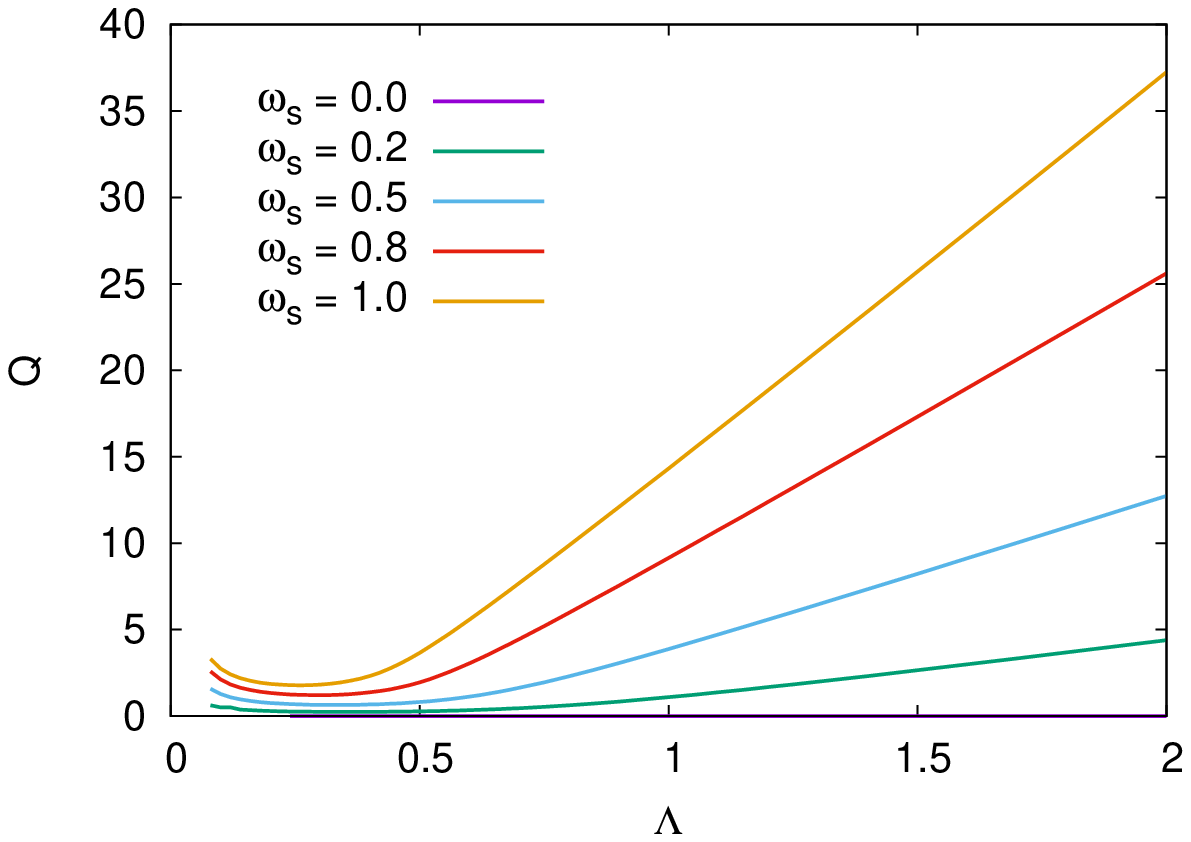}
\label{Fig1d}
}
}
\mbox{\hspace{0.2cm}
\subfigure[][]{\hspace{-1.0cm}
\includegraphics[height=.20\textheight, angle =0]{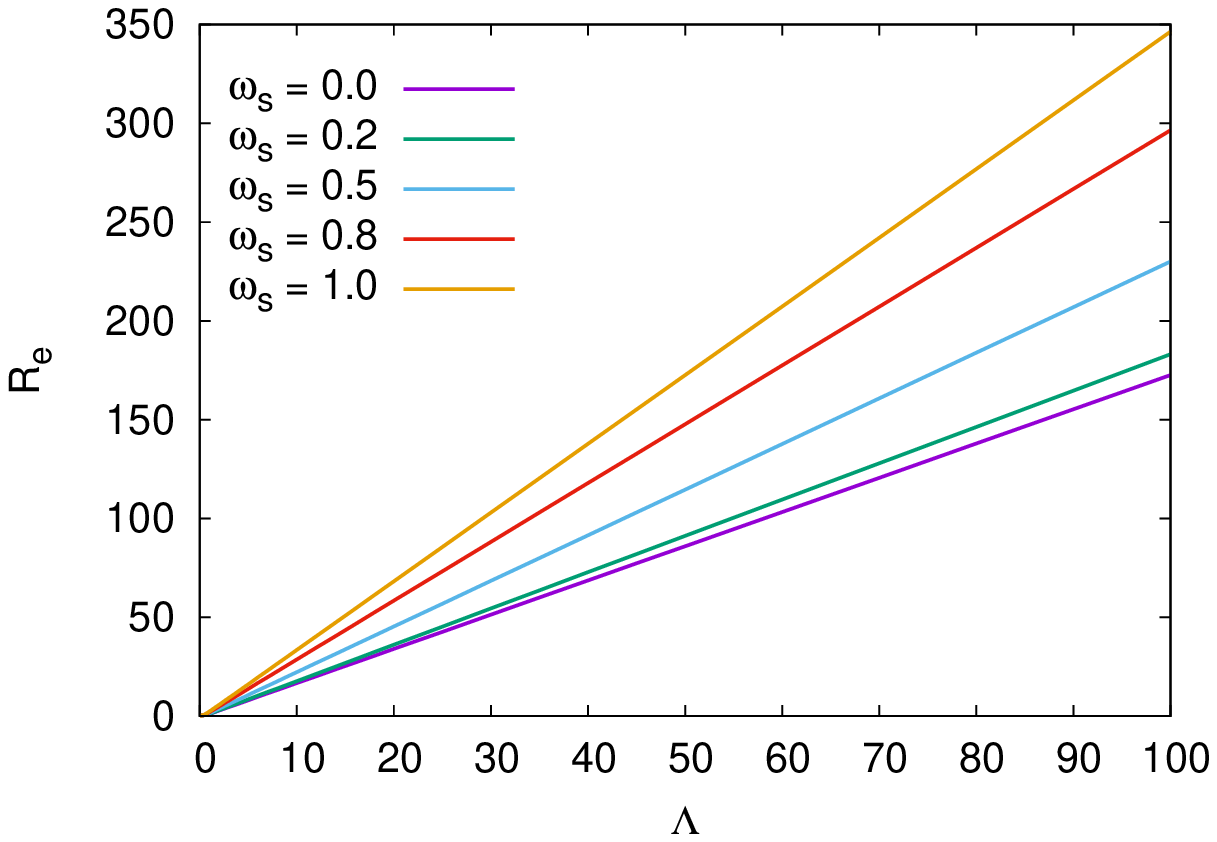}
\label{Fig1e}
}
\subfigure[][]{\hspace{-0.5cm}
\includegraphics[height=.20\textheight, angle =0]{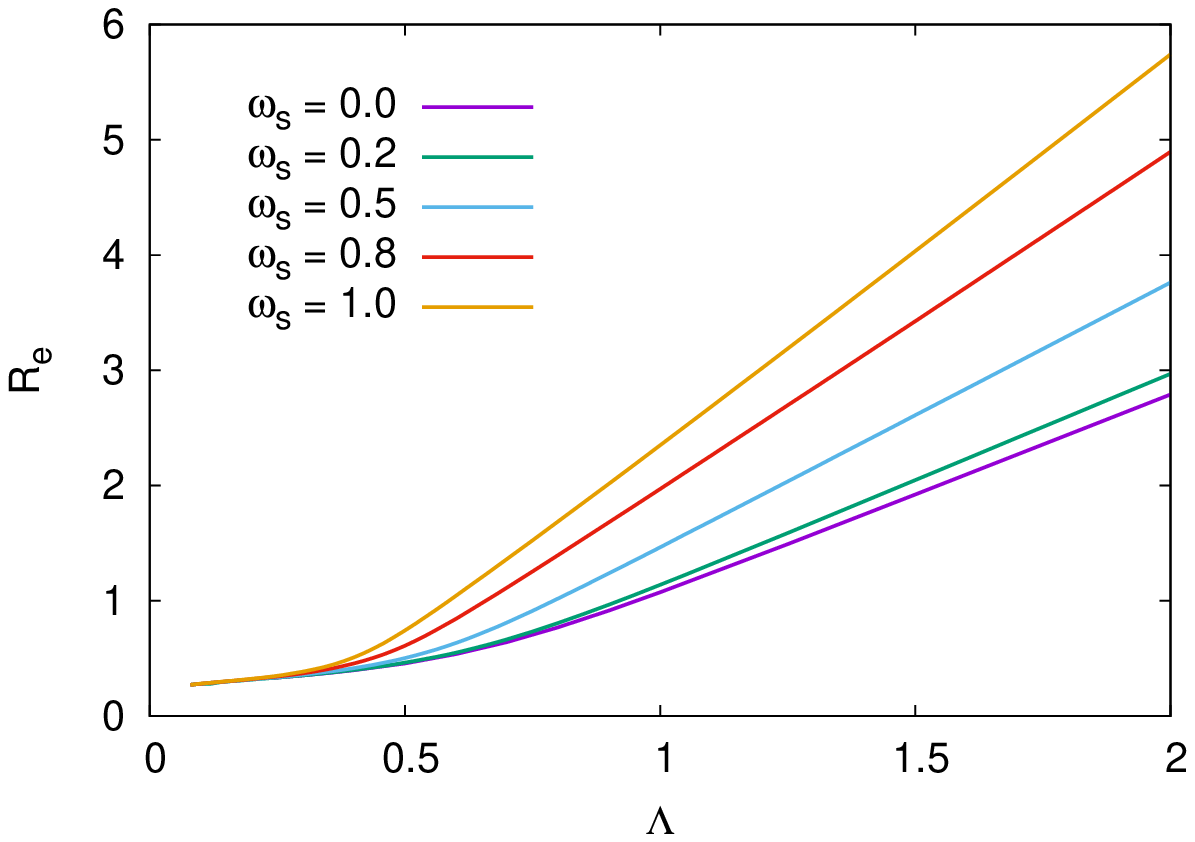}
\label{Fig1f}
}
}
\mbox{\hspace{0.2cm}
\subfigure[][]{\hspace{-1.0cm}
\includegraphics[height=.20\textheight, angle =0]{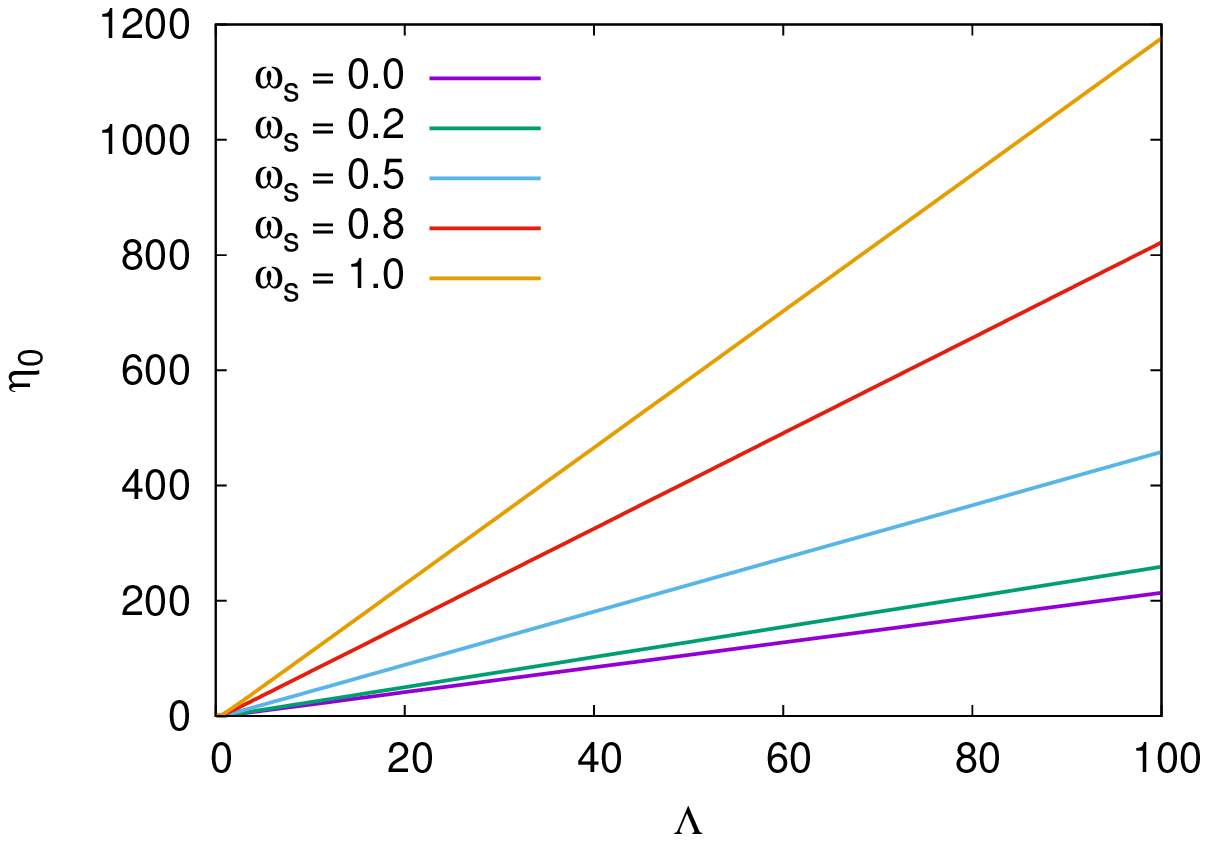}
\label{Fig1g}
}
\subfigure[][]{\hspace{-0.5cm}
\includegraphics[height=.20\textheight, angle =0]{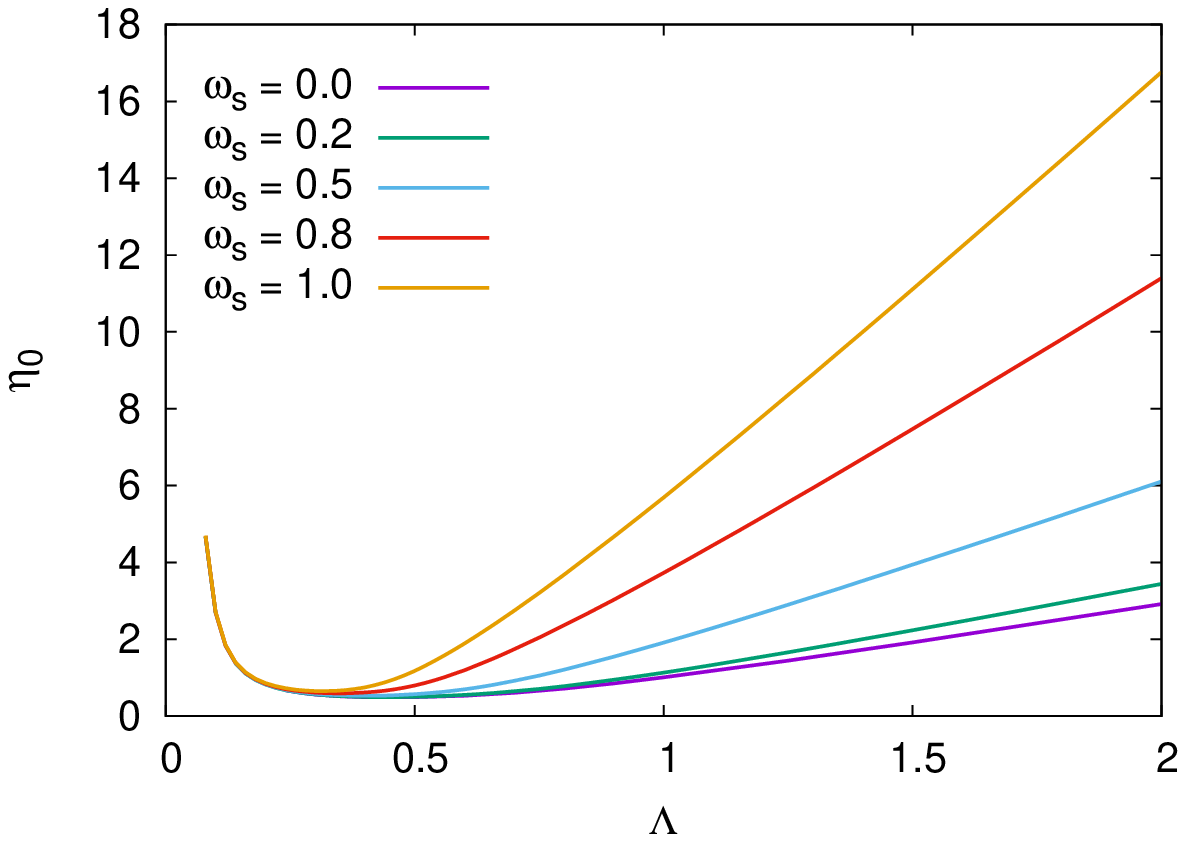}
\label{Fig1h}
}
}
\end{center}
\vspace{-0.5cm}
\caption{Properties of non-rotating wormhole solutions
($n=0$) versus the coupling strength $\Lambda$
($0.08 \leq \Lambda \leq 100$)
for several values of the boson frequency $\omega_s$:
(a) mass $M$;
(b) zoom of (a);
(c) particle nummber $Q$;
(d) zoom of (c);
(e) circumferential radius $R_e$;
(f) zoom of (e);
(g) throat parameter $\eta_0$;
(h) zoom of (g);
\label{Fig1}
}
\end{figure}

Non-rotating solutions appear to exist for all values of the
coupling strength $\Lambda > 0$,
and in the full interval of the boson frequency $0 \leq \omega_s \leq 1$.
In Fig.\ref{Fig1} we show  the mass $M$, particle number $Q$, circumferential radius $R_e$ and
throat parameter $\eta_0$ as a function of the coupling strength $\Lambda$ for several values of $\omega_s$.
For the limiting cases $\omega_s=0$ and $\omega_s=1$ 
the mass, particle number and throat radius
assume finite values (except for $Q=0$ when $\omega_s=0$). 
This is in contrast to solutions with an ordinary complex scalar field
describing boson stars or wormholes immersed inside bosonic matter.
We emphasize that the solutions become static when $\omega_s=0$,
since the time-dependence of the phantom field then disappears.

In the limit $\Lambda \to \infty$
the particle number increases linearly with $\Lambda$. 
The mass increases (decreases) linearly
for $\omega_s>0.5$ ($\omega_s<0.5$) 
and assumes a finite value for $\omega_s=0.5$.
The throat radius increases linearly with $\Lambda$.
For small values of $\Lambda$ the particle number 
possesses a minimum at some $\Lambda_{\rm min}$
(as long as $\omega_s\neq 0$) 
and increases with decreasing $\Lambda$ for $\Lambda<\Lambda_{\rm min}$.
The mass possesses a maximum only if $\omega_s > 0.5$. 
The throat radius increases with increasing $\Lambda$ for large values of $\Lambda$.
For small values of $\Lambda$, the throat radius tends to some finite value $R_e^{(0)}$
where $R_e^{(0)}$ is independent of $\omega_s$.
Like the radius, the throat parameter $\eta_0$ increases linearly with increasing $\Lambda$ 
for large values of $\Lambda$ and is independent of $\omega_s$ when $\Lambda$ becomes small.
However, in this case $\eta_0$ increases as $\Lambda$ decreases.

\subsection{Rotating solutions}

We now turn to the rotating wormhole solutions.
We note that in all the rotating wormhole solutions
considered here it is the rotation of the phantom field
which induces the rotation of the spacetime.
This is very much in contrast to the rotating Ellis wormholes,
where the rotation is imposed via the (asymmetric) boundary conditions
\cite{Kleihaus:2014dla,Chew:2016epf}.

\begin{figure}[p!]
\mbox{\hspace{0.2cm}
\subfigure[][]{\hspace{-1.0cm}
\includegraphics[height=.30\textheight, angle =0]{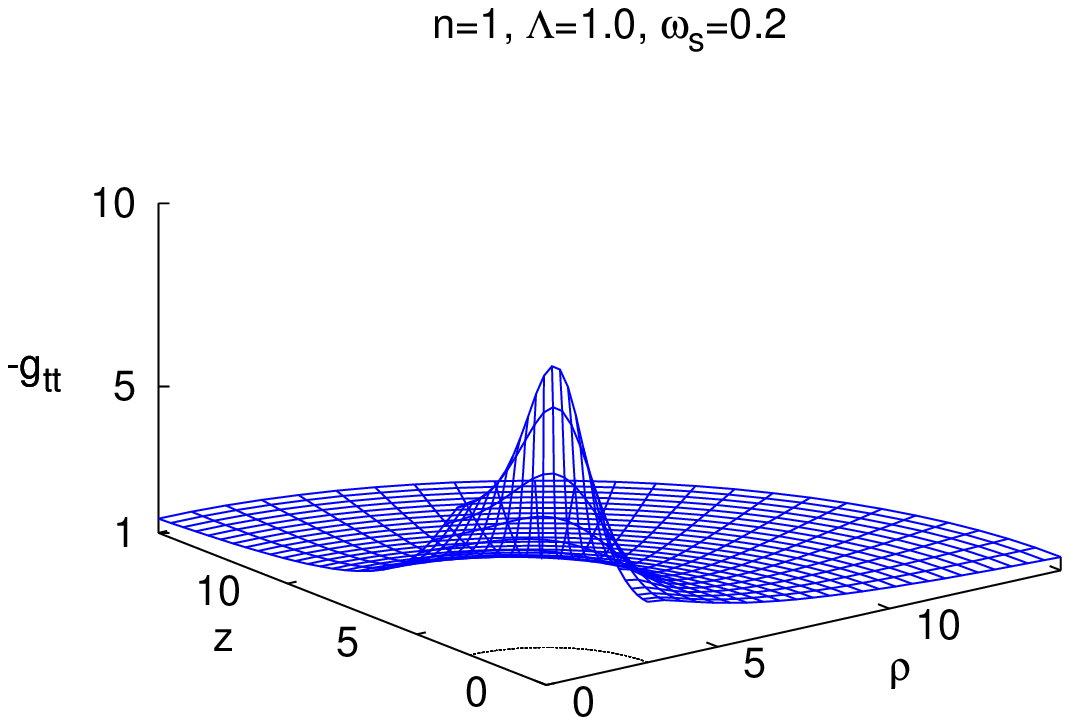}
\label{Fig2a}
}
\subfigure[][]{\hspace{-0.5cm}
\includegraphics[height=.30\textheight, angle =0]{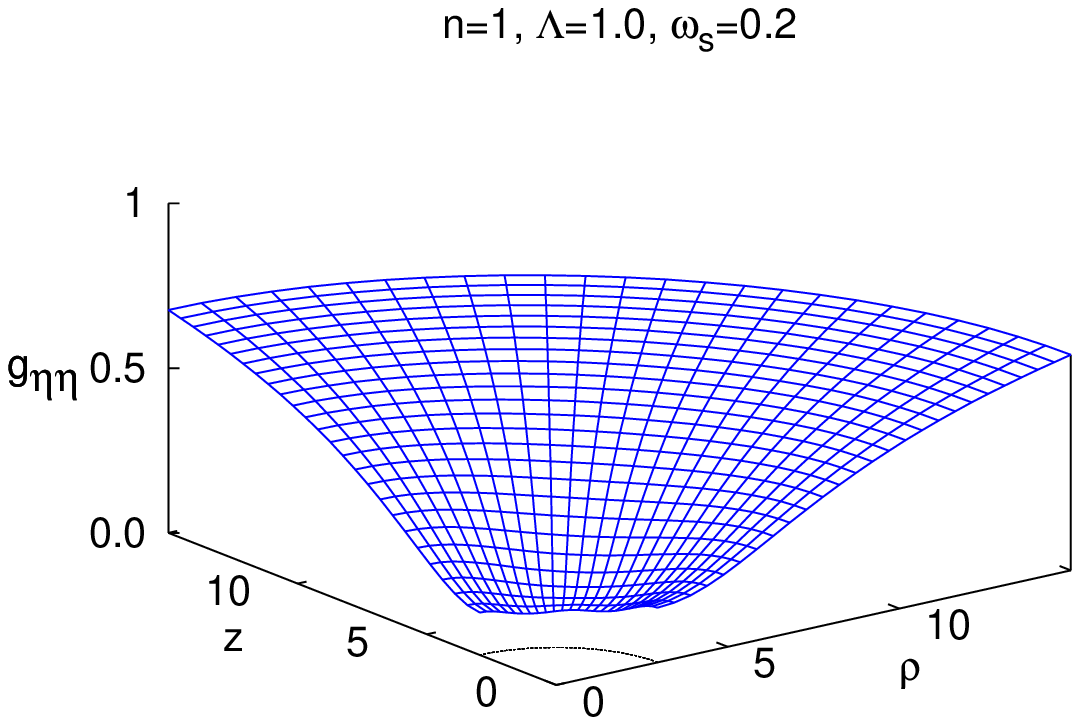}
\label{Fig2b}
}
}
\mbox{\hspace{0.2cm}
\subfigure[][]{\hspace{-1.0cm}
\includegraphics[height=.30\textheight, angle =0]{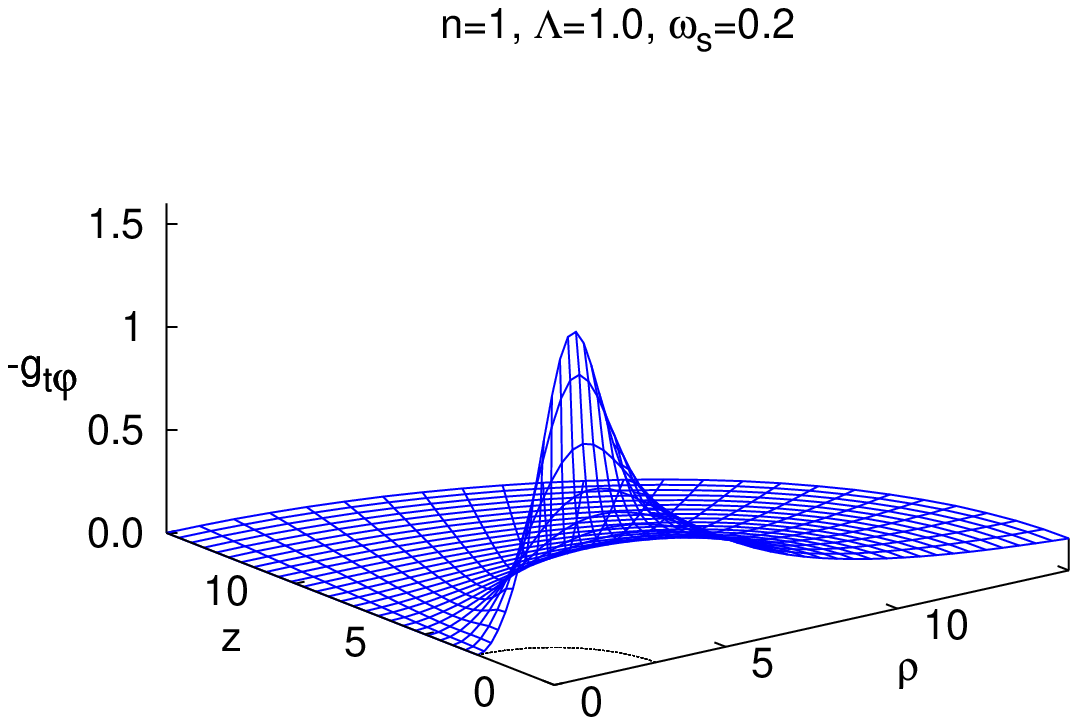}
\label{Fig2c}
}
\subfigure[][]{\hspace{-0.5cm}
\includegraphics[height=.30\textheight, angle =0]{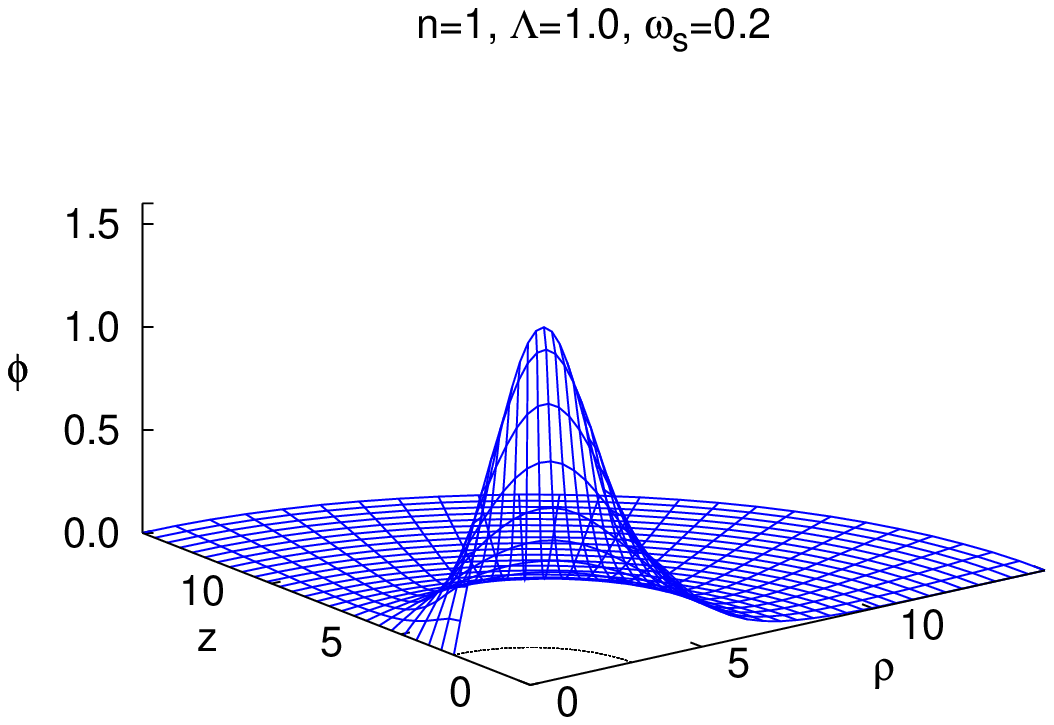}
\label{Fig2d}
}
}
\mbox{\hspace{0.2cm}
\subfigure[][]{\hspace{-1.0cm}
\includegraphics[height=.30\textheight, angle =0]{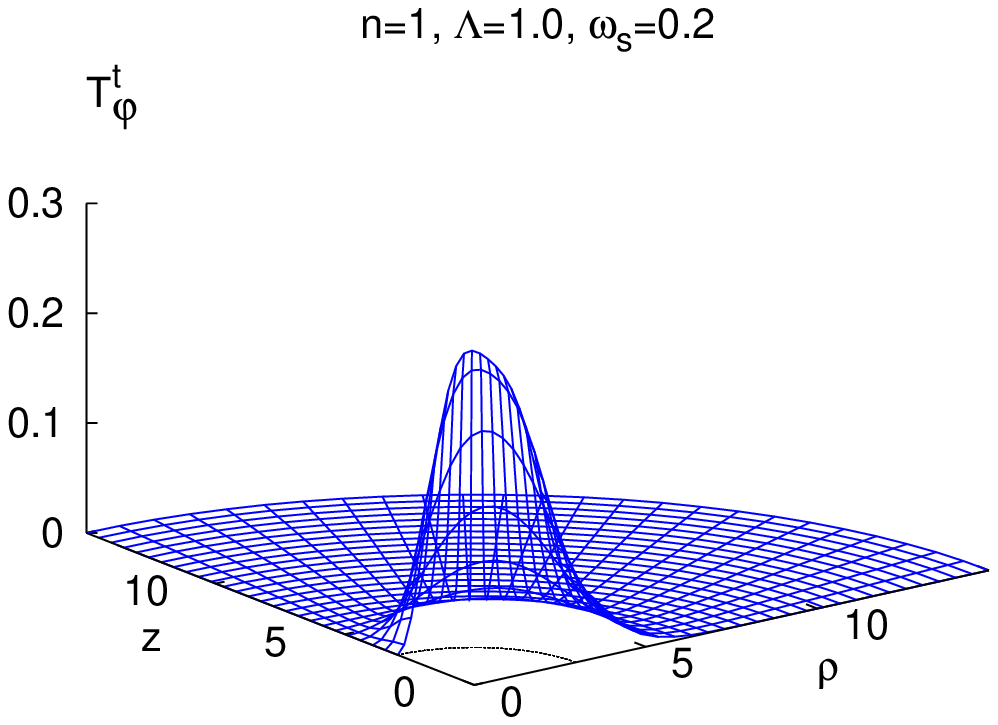}
\label{Fig2e}
}
\subfigure[][]{\hspace{-0.5cm}
\includegraphics[height=.30\textheight, angle =0]{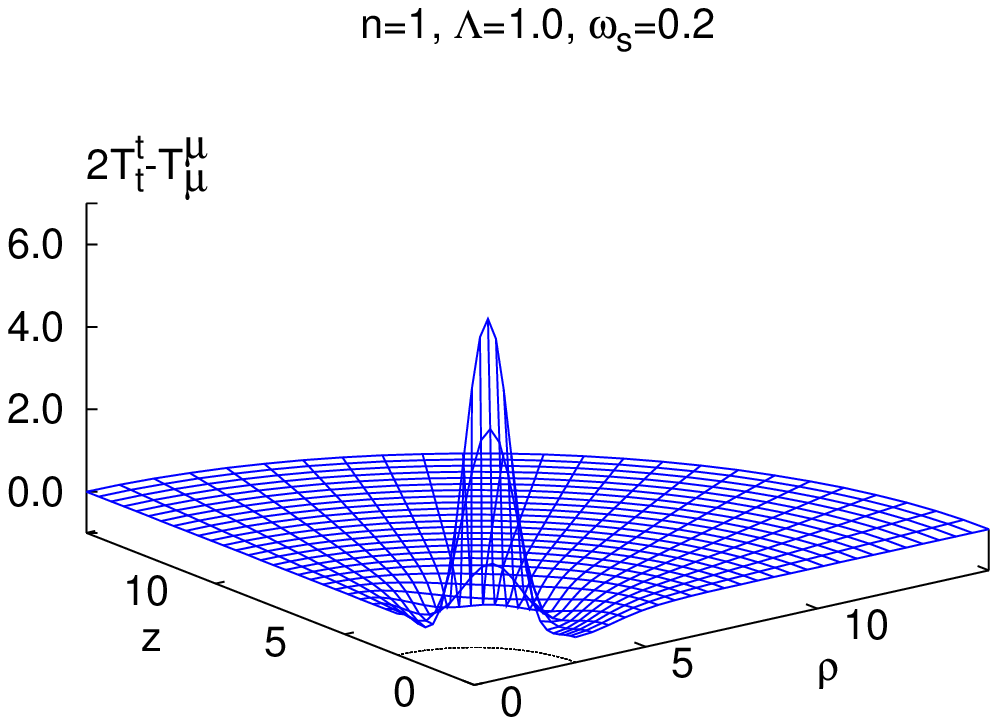}
\label{Fig2f}
}
}
\caption{Rotating wormhole solution
($n=1$, $\Lambda=1$, $\omega_{\rm s}=0.2$):
(a) metric component $-g_{tt}$;
(b) metric component $g_{\eta\eta}$;
(c) metric component $-g_{t\varphi}$;
(d) phantom field function $\phi$;
(e) angular momentum density $T^t_\varphi$;
(f) Komar mass density $2T^t_t - T^\mu_\mu$
in cylindrical coordinates based on
an isotropic radial coordinate.
The circle in the $\rho z$-plane indicates the location of the throat.
\label{Fig2}
}
\end{figure}

\subsubsection{Numerical scheme}

We have constructed a large number of wormhole solutions
with winding number $n=1$ numerically,
covering the boson frequency interval $0 \leq \omega_s \leq  1$,
and the interval for the self-interaction strength
$0.7 \leq \Lambda < 2.0$.
For values of $\Lambda$ outside this interval
the numerical errors have increased too much, to consider
the solutions any longer as being fully reliable.

To solve the system of coupled partial differential equations
we have employed the routine FIDISOL/CADSOL \cite{schonauer:1989},
which is a finite difference solver based on a Newton-Raphson scheme.
We have introduced a compactified coordinate $x=\arctan(\eta)$,
to obtain a finite coordinate patch. Then we have chosen
a non-equidistant grid with typically $130 \times 50$ grid points
in radial ($x$) and angular ($\theta$) direction.

For given values of $\omega_s$ and $\Lambda$ 
we have then adjusted the parameter $\eta_0$ such that 
the constraint Eq.~(\ref{constrd}) vanishes
(to a given accuracy).
In particular, we have introduced the $L_2$ norm of the constraint
\begin{equation}
D(\eta_0) = \sqrt{\sum_{i,j} d^2(x_i,\theta_j)} \ ,
\label{defD}
\end{equation}
where the sum is over all inner gridpoints,
and have determined the minimum of $D$ with respect to $\eta_0$. 
Typical values of $D$ at the minimum have been 
in the range $10^{-4}$-$10^{-3}$, which is 
comparable with the $L_2$ norm of the solutions of the PDEs.

\subsubsection{Solutions}

We begin our discussion by exhibiting in Fig.~\ref{Fig2}
the metric and the phantom field of a typical solution,
where we have chosen the parameters
$n=1$, $\Lambda=1$ and $\omega_{\rm s}=0.2$.
The coordinates employed in the figure are 
cylindrical coordinates based on an isotropic radial
coordinate
\begin{equation}
\rho = r\sin\theta  
\ , \ \ \
z= r\cos\theta
\ , \ \ \
\eta= r_0\left(\frac{r}{r_0}-\frac{r_0}{r}\right) 
\ , \ \ \
r_0 = \frac{\eta_0}{2}
\ .
\label{coor}
\end{equation}
The figure shows the metric component $-g_{tt}$,
the metric component $g_{\eta\eta}$,
the metric component $-g_{t\varphi}$,
the phantom field function $\phi$. 
Also shown are 
the angular momentum density $T^t_\varphi$,
and the Komar mass density $2T^t_t - T^\mu_\mu$, appearing in the
Komar integrals Eqs.~(\ref{intj}), resp.~(\ref{intm}).

The metric functions $g_{tt}$ and $g_{\eta\eta}$
differ most pronouncedly from their
asymptotic values at the throat.
Here it comes as a surprise that the maximal value of
$|g_{tt}|$ does not arise at the throat in the equatorial plane.
Therefore the maximum is attained on two rings
on the throat, one in the upper hemisphere
and one in the lower hemisphere
associated with angles $\theta_m$ and $\pi - \theta_m$, respectively.
The metric function $-g_{t\varphi}$ exhibits
a ring of saddle points on the throat in the equatorial plane and
maxima located also roughly
at $\theta_m$ and $\pi - \theta_m$.

Like $g_{tt}$ and $g_{\eta\eta}$,
the phantom field function assumes its largest deviations
from its vacuum value at the throat,
and again the angles $\theta_m$ and $\pi - \theta_m$
indicate the rings of maxima.
Therefore the rings of maxima of the angular momentum density
and the Komar mass density found in the vicinity of these angles
are to be expected.

Thus the picture we find for the metric, the phantom field and the
stress energy tensor differs fundamentally from 
the one encountered in boson stars or wormholes immersed
in ordinary rotating bosonic matter. There the matter is concentrated
in a torus centered in the equatorial plane.
Here the phantom matter is concentrated in two tori,
located symmetrically with respect to the equatorial plane.
In particular, we here find two tori for a complex field,
that is symmetric with respect to reflection at $\theta=\pi/2$.
Recall that double tori configurations arise for boson stars
only when the complex field is antisymmetric 
with respect to reflection at $\theta=\pi/2$, i.e., for
negative parity configurations \cite{Kleihaus:2007vk}.

Having noted this fundamental difference in the configurations obtained
with an ordinary complex scalar field and a complex phantom field,
let us address the dependence on the parameters. 
In fact, we observe that this fundamental difference remains, as $\omega_s$
and $\Lambda$ are varied, and only the value of $\theta_m$ shifts.

Another fundamental difference to the known boson stars
and wormholes immersed in ordinary rotating bosonic matter
is the presence of $\omega_s=0$ solutions.
These solutions are perfectly regular,
and they represent static deformed wormholes.
The deformation is induced by the $\varphi$-dependence of the 
phantom field, since $n\ne 0$.

Concerning the presence of ergoregions,
where $g_{tt}>0$, we note
that no ergoregions have emerged for the solutions 
we have studied so far.
Also this is in constrast to boson stars
and wormholes immersed in ordinary rotating bosonic matter,
since those solutions are known to feature ergoregions
for (sufficiently) fast rotation \cite{Kleihaus:2007vk,Hoffmann:2017vkf,Hoffmann:2018oml}.

\subsubsection{Global charges}

\begin{figure}[t!]
\begin{center}
\mbox{\hspace{0.2cm}
\subfigure[][]{\hspace{-1.0cm}
\includegraphics[height=.25\textheight, angle =0]{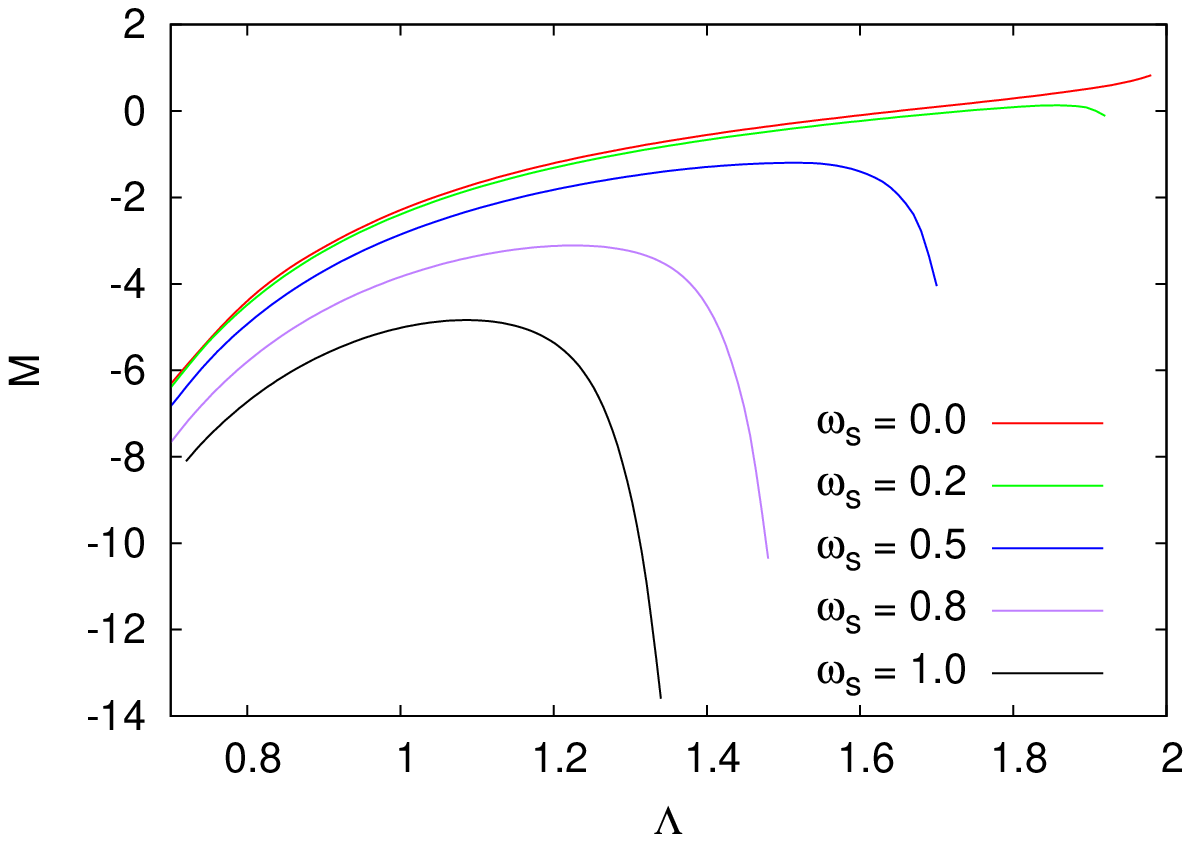}
\label{Fig3a}
}
\subfigure[][]{\hspace{-0.5cm}
\includegraphics[height=.25\textheight, angle =0]{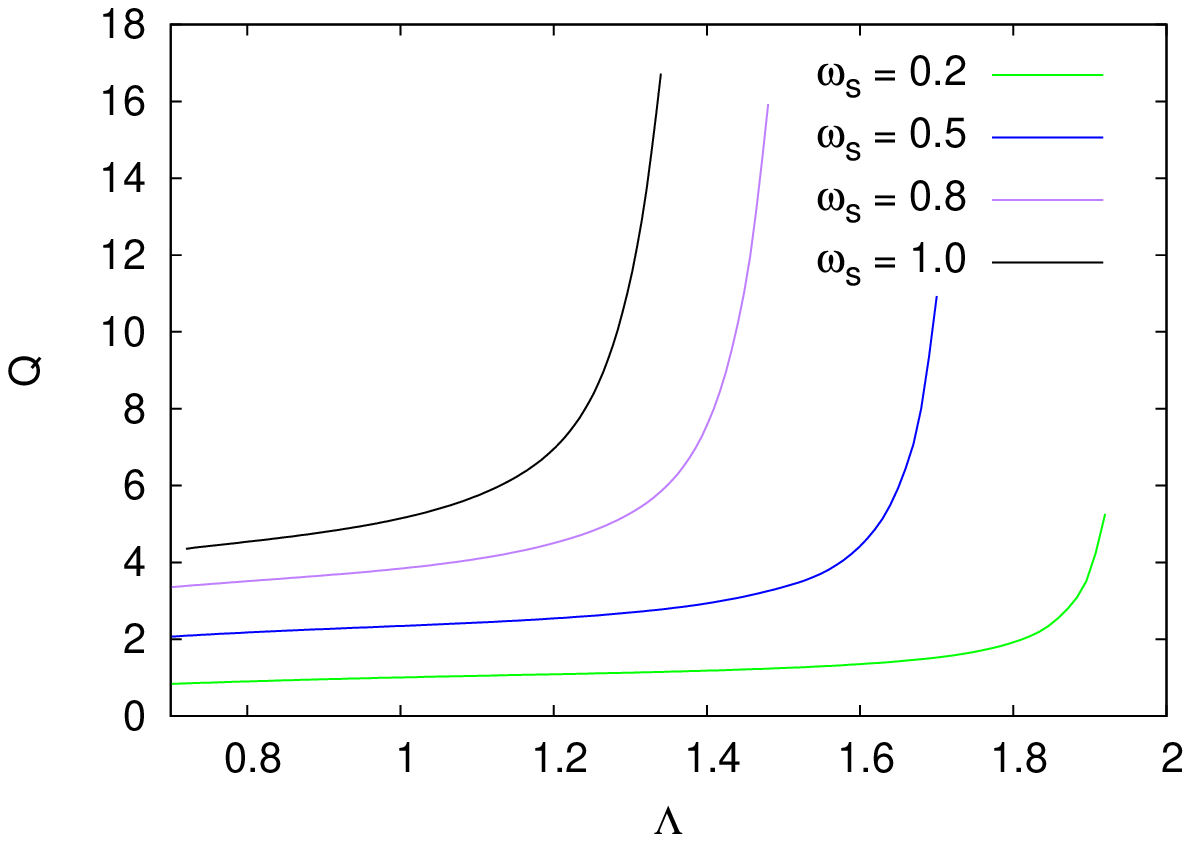}
\label{Fig3b}
}
}
\end{center}
\vspace{-0.5cm}
\caption{Global charges of rotating solutions ($n=1$):
(a) mass $M$ versus coupling strength $\Lambda$
for several values of the boson frequency $\omega_s$; 
(b) same as (a) for the particle number $Q$.
\label{Fig3}
}
\end{figure}

Let us now turn to the global charges of the 
rotating wormhole solutions and their dependence on the parameters.
In Fig.\ref{Fig3} we show the mass $M$ and the particle number $Q$
versus the coupling strength $\Lambda$
for several values of the boson frequency $\omega_s$,
including the limits $\omega_s=0.0$ and $1.0$.
Recall that here the angular momentum $J$ agrees with the particle number
since $n=1$.

We note that for $\omega_s=0$ static solutions
arise, which carry neither particle number nor angular momentum,
but possess the highest mass for a given value of the
coupling strength.
For $\omega_s=0$ the mass increases with increasing $\Lambda$,
in the range of $\Lambda$ considered.

For finite values of $\omega_s$, the solutions rotate
and possess a finite angular momentum.
Whereas for the smaller values of $\Lambda$ the mass first increases 
and then decreases, the angular momentum always increases
with $\Lambda$ up to a certain point, where
the numerical accuracy deteriorates,
and we do not depict the solutions any longer in the figure.

When the solutions are still sufficiently accurate, we note
that the larger $\omega_s$ the smaller the mass
for a given $\Lambda$.
At the same time, the larger $\omega_s$
the earlier the rapid decrease of the mass sets in,
and the earlier numerical accuracy is lost.
Note that the rapid decrease of the mass goes along with
a rapid increase of the particle number and angular momentum.

\subsubsection{Geometrical  properties}

\begin{figure}[t!]
\begin{center}
\mbox{\hspace{0.2cm}
\subfigure[][]{\hspace{-1.0cm}
\includegraphics[height=.25\textheight, angle =0]{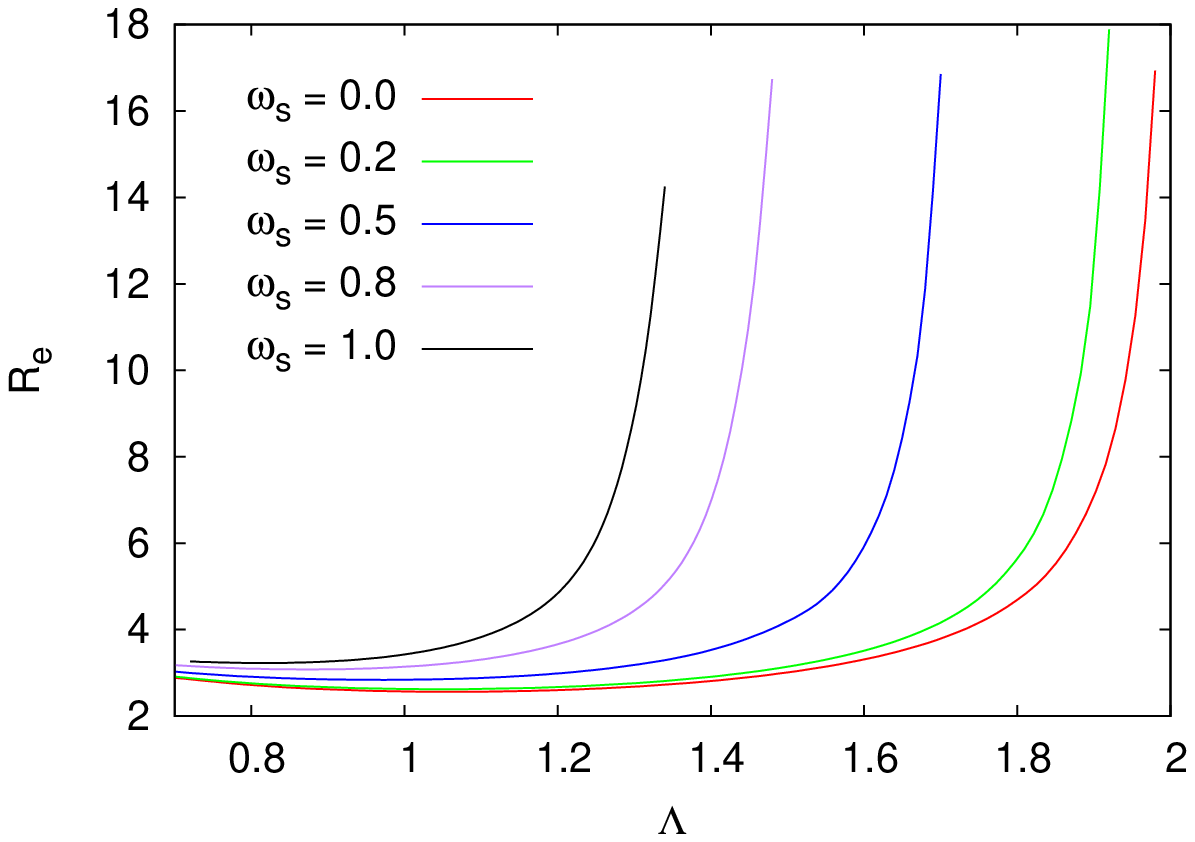}
\label{Fig4a}
}
\subfigure[][]{\hspace{-0.5cm}
\includegraphics[height=.25\textheight, angle =0]{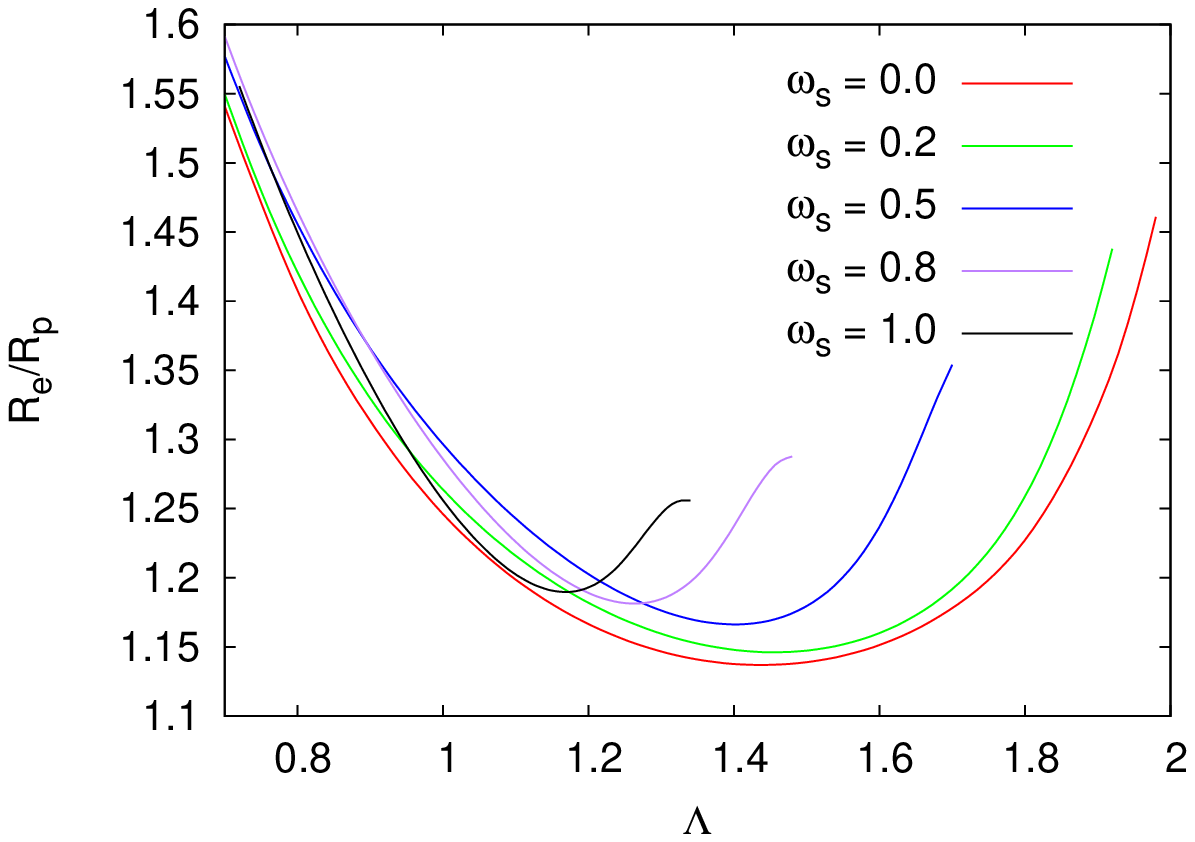}
\label{Fig4b}
}
}
\mbox{\hspace{0.2cm}
\subfigure[][]{\hspace{-1.0cm}
\includegraphics[height=.25\textheight, angle =0]{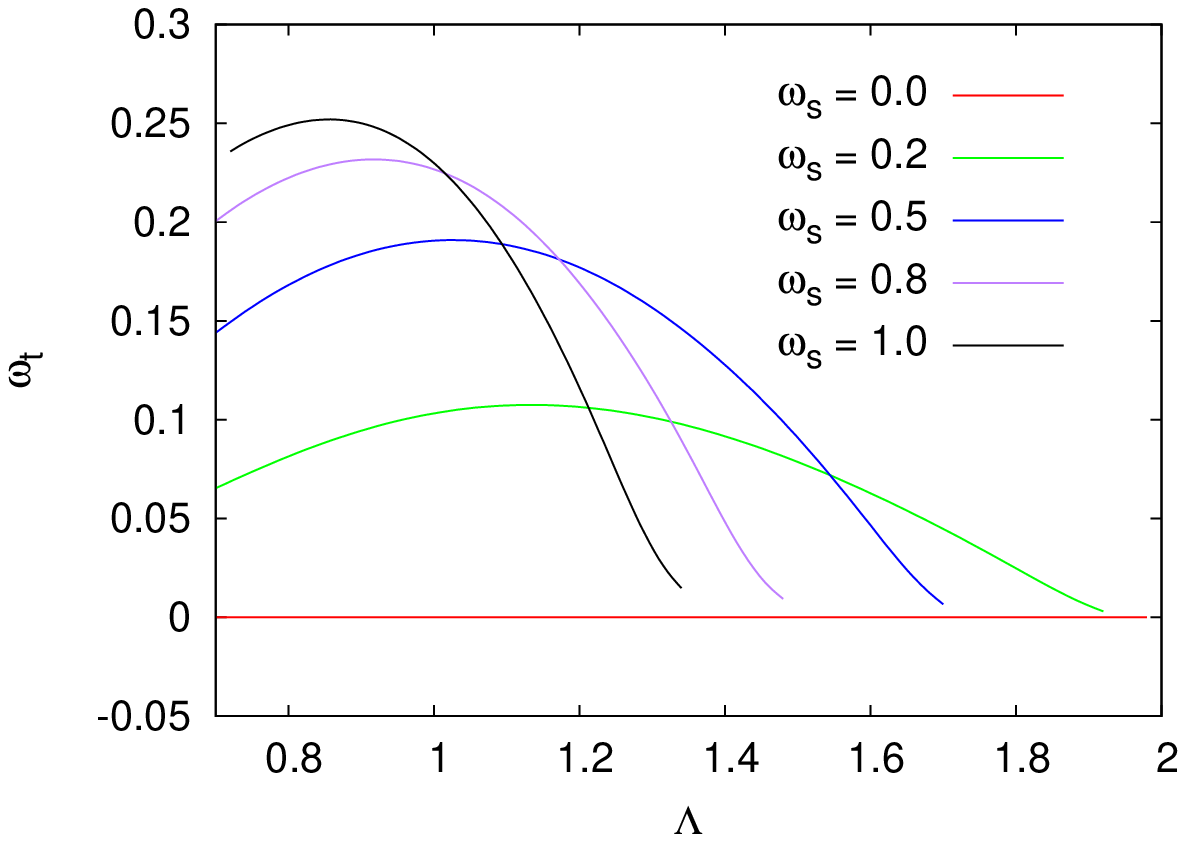}
\label{Fig4c}
}
\subfigure[][]{\hspace{-0.5cm}
\includegraphics[height=.25\textheight, angle =0]{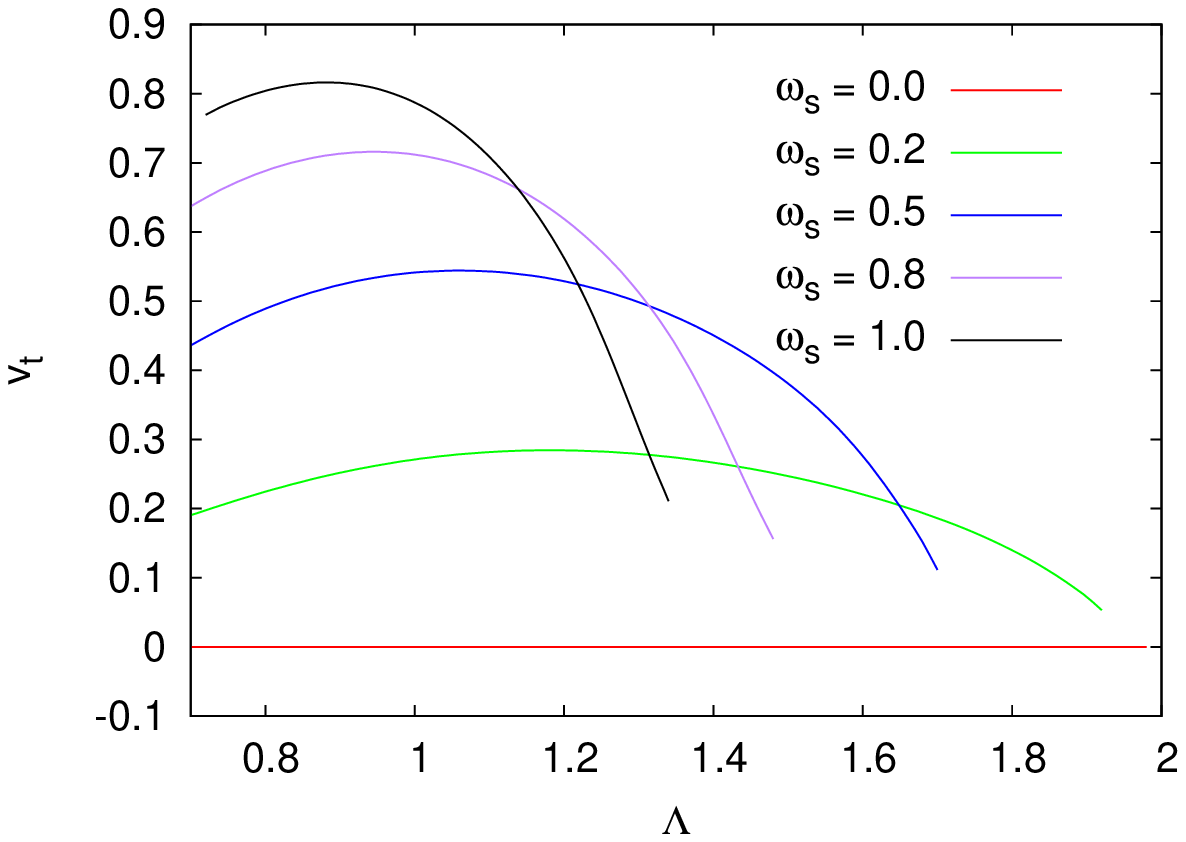}
\label{Fig4d}
}
}
%
\end{center}
\vspace{-0.5cm}
\caption{Geometric properties of solutions with $n=1$
versus the coupling strength $\Lambda$
for several values of the boson frequency $\omega_s$:
(a) the circumferential radius  $R_{\rm e}$ of the throat 
in the equatorial plane;
(b) the ratio of circumferential equatorial
and polar radii $R_{\rm e}/R_{\rm p}$;
(c) the rotational frequency $\omega_{\rm t}$
of the throat in the equatorial plane;
(d) the rotational velocity $v_{\rm t}$
of the throat in the equatorial plane.
\label{Fig4}
}
\end{figure}

Let us next address the geometrical properties of the rotating solutions,
focusing mostly on the equatorial plane.
We present in Fig.\ref{Fig4} the dependence on the coupling strength
$\Lambda$ for several physically interesting quantities:
the circumferential radius  $R_{\rm e}$ (a),
the ratio of the circumferential equatorial and polar radii
$R_{\rm e}/R_{\rm p}$ (b),
the rotational frequency of the throat $\omega_{\rm t}$ 
in the equatorial plane (c),
and the rotational velocity of the throat $v_{\rm t}$ 
in the equatorial plane (d),
for fixed values of the boson frequency $\omega_s$.

For a fixed boson frequency,
the circumferential radius in the equatorial plane 
first changes slowly for the smaller values of $\Lambda$, but then
exhibits a strong growth in analogy to the strong
growth of the angular momentum and the strong decrease
of the mass.
Likewise, for a given $\Lambda$, the circumferential radius
is the larger, the larger the boson frequency.
Note that wormholes with multiple throats and equators have not been found
in the parameter space considered.

The ratio of the polar and equatorial
circumferential radii is expected to give some
insight into the deformation of the throat,
since for a spherical wormhole throat the ratio would be unity.
Here the figure shows that this ratio starts from 
a large deformation at the smallest $\Lambda$ considered,
where it does not vary much with the boson frequency.
Then the deformation decreases, reaches a minimum and increases
again with increasing $\Lambda$.

The rotational frequency of the throat 
in the equatorial plance vanishes for 
$\omega_s=0$, and its maximum increases with increasing $\omega_s$,
shifting to smaller values of $\Lambda$.
When the throat rapidly increases its size
with increasing $\Lambda$ the rotational frequency of the throat
in the equatorial plane tends towards zero.
The rotational velocity of the throat
in the equatorial plane follows this behavior to some extent,
since it is defined by $v_t = \omega_t R_{\rm e}$.

\begin{figure}[t!]
\begin{center}
\mbox{\hspace{0.2cm}
\subfigure[][]{\hspace{-1.0cm}
\includegraphics[height=.25\textheight, angle =0]{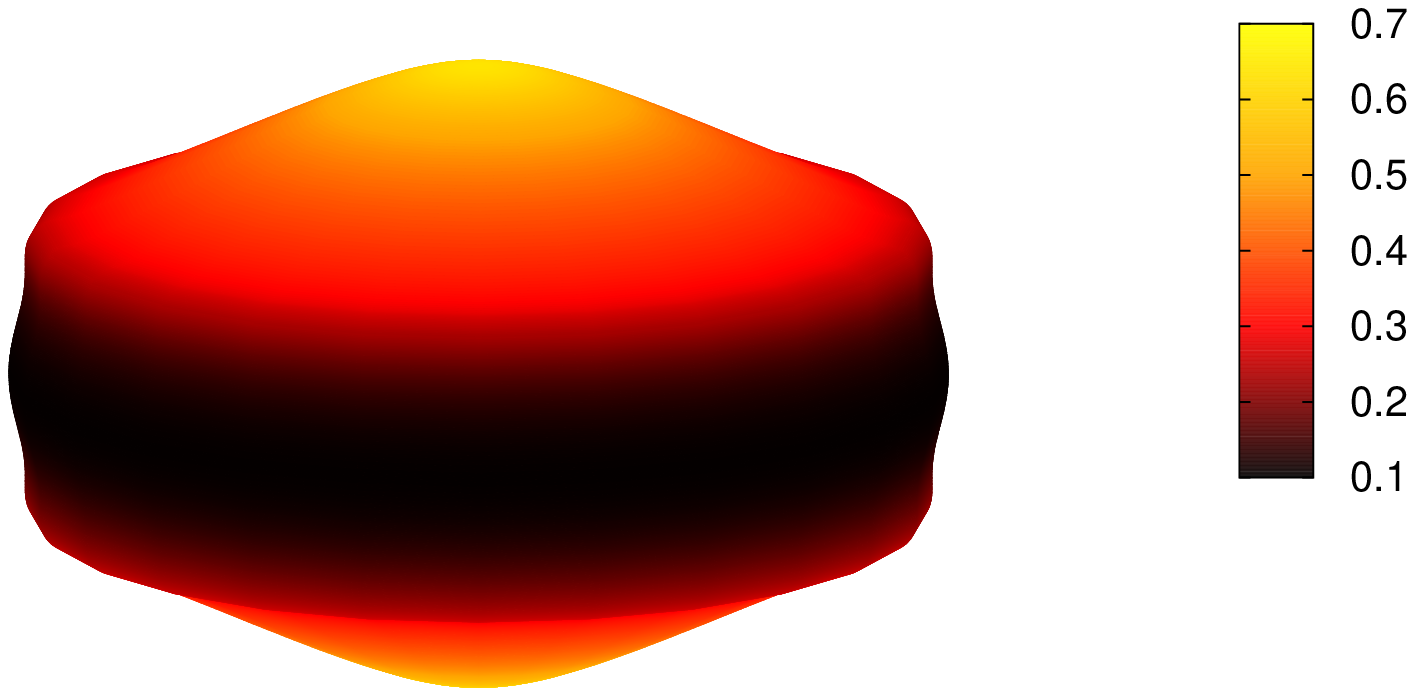}
\label{Fig5a}
}
\subfigure[][]{\hspace{-0.5cm}
\includegraphics[height=.25\textheight, angle =0]{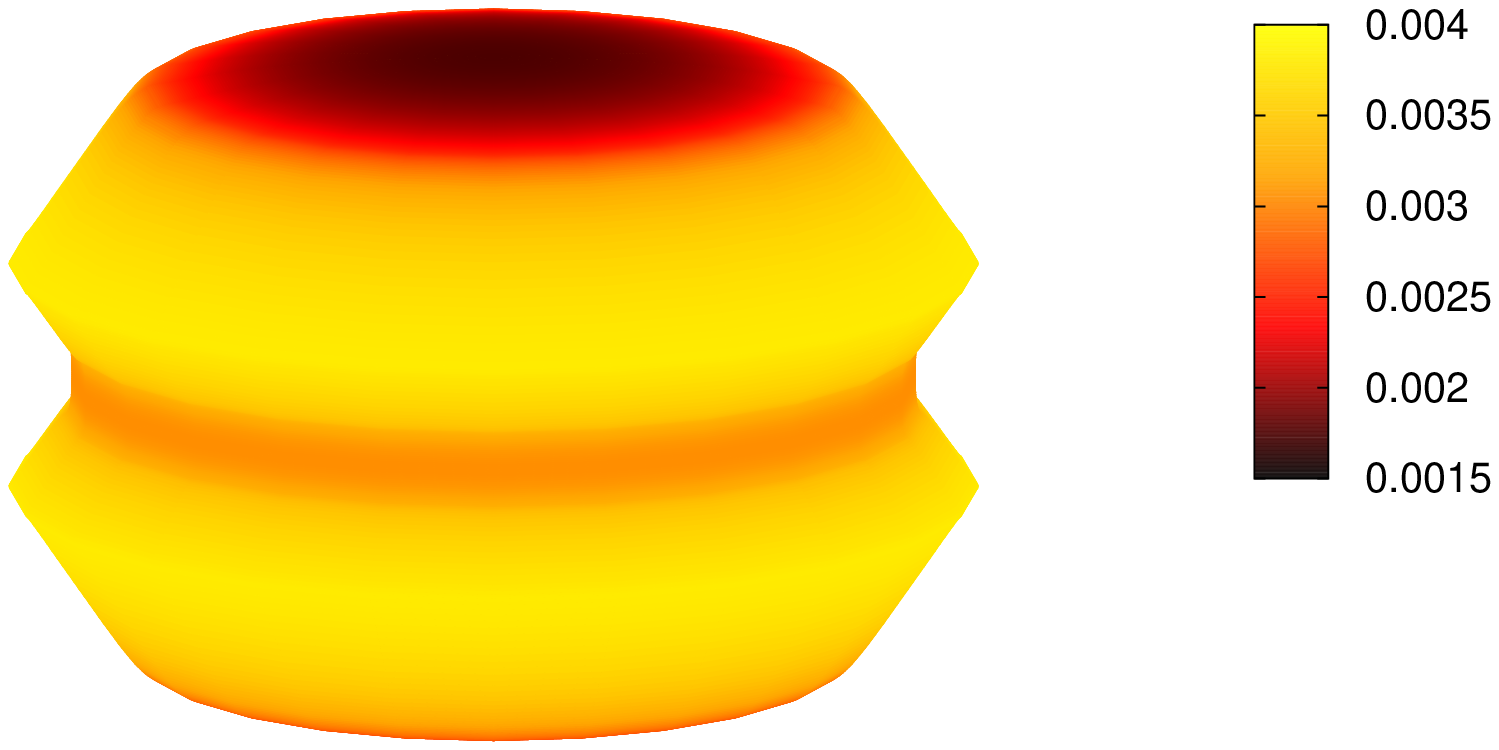}
\label{Fig5b}
}
}
\mbox{\hspace{0.2cm}
\subfigure[][]{\hspace{-1.0cm}
\includegraphics[height=.25\textheight, angle =0]{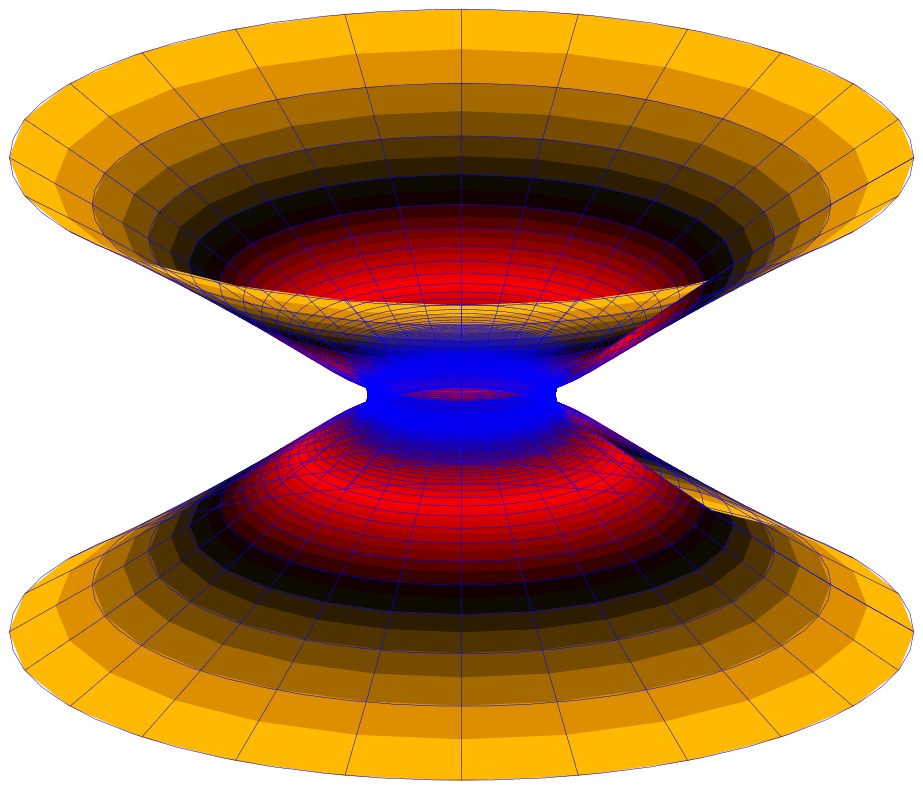}
\label{Fig5c}
}
\subfigure[][]{\hspace{-0.5cm}
\includegraphics[height=.25\textheight, angle =0]{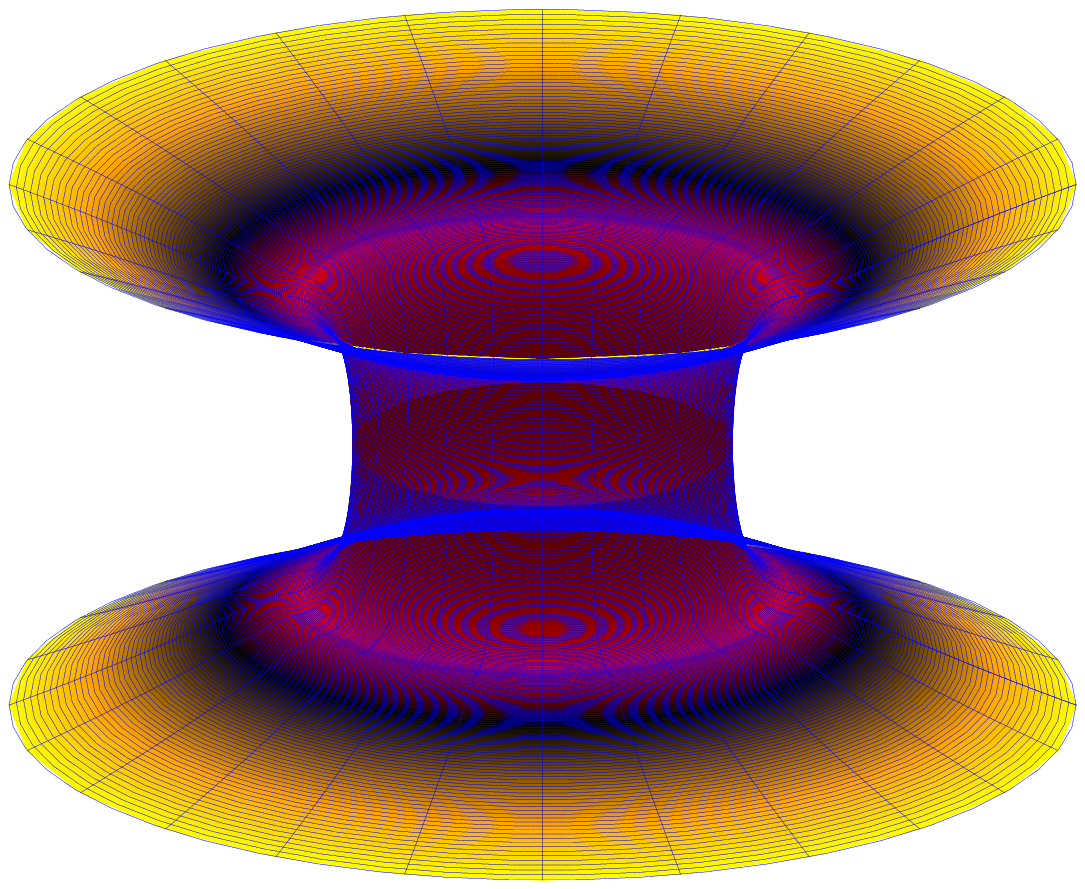}
\label{Fig5d}
}
}
\end{center}
\vspace{-0.5cm}
\caption{Rotating solutions ($n=1$):
(a) angular frequency on the embedded throat
($\omega_s=0.2$, $\Lambda=1$);
(b) same as (a) for $\Lambda=1.92$;
(c) embedding diagram of the throat in the equatorial plane
($\omega_s=0.2$, $\Lambda=1$);
(d) same as (c) for $\Lambda=1.92$.
}
\label{Fig5}
\end{figure}

The figure nicely demonstrates
that the rotation of the phantom field indeed induces a rotation
of the throat and consequently a rotation of the full spacetime.
However, this rotation is not uniform on the throat
but depends on the polar angle.
We exhibit in Fig.~\ref{Fig5} the rotational frequency on
the embedded throat for two solutions ($n=1$, $\omega_s=0.2$, $\Lambda= 1$
and $\Lambda= 1.92$). We note that the embedded throat deviates strongly
from a sphere. Moreover, the embedding is partly pseudo-euclidean.

To get further insight into the geometry of these wormholes 
we exhibit in Fig.~\ref{Fig5} also isometric embeddings
of the equatorial plane for the same two solutions.
We see that for the larger self-coupling parameter the 
throat is more prolonged. Note that the embedding is euclidean close 
to the throat, but pseudo-euclidean otherwise.

\begin{figure}[t!]
\begin{center}
\mbox{\hspace{0.2cm}
\subfigure[][]{\hspace{-1.0cm}
\includegraphics[height=.20\textheight, angle =0]{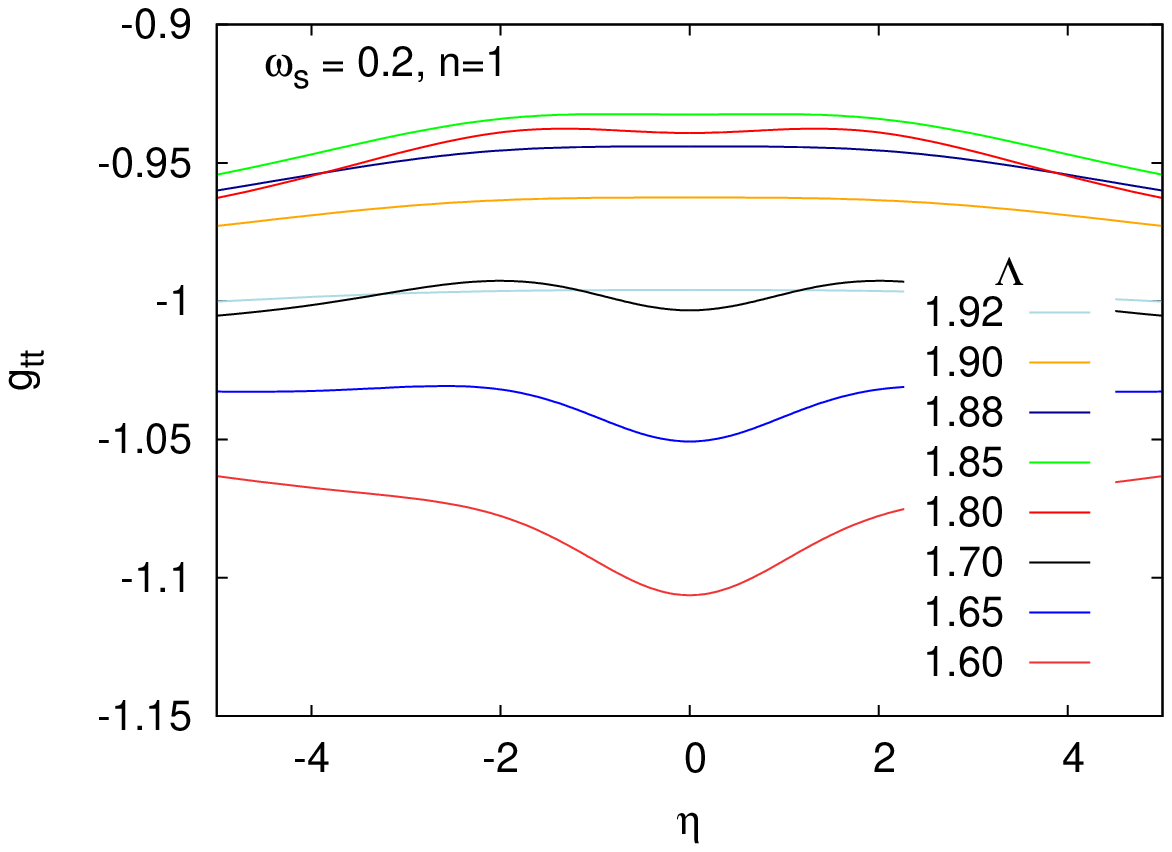}
\label{Fig6a}
}
}
\end{center}
\vspace{-0.5cm}
\caption{Static orbits:
We show the metric component $g_{tt}$ in the equatorial plane
as a function of $\eta$ for $\omega_s=0.2$ and several values of
$\Lambda$.
\label{Fig6}
}
\end{figure}
Let us finally consider the static orbits. 
As examples we consider rotating wormhole solutions with $\omega_s=0.2$.
In Fig.\ref{Fig6} we show the metric component $g_{tt}$  in the equatorial plane
as a function of $\eta$. 
Stable static orbits exist only for values of $\Lambda$ larger than
$\approx 1.6$. If  $\Lambda> 1.85$  a single stable static orbit is 
located at the throat. For smaller values of $\Lambda$ two stable static 
orbits exist, located symmetrically to the left and to the right of the throat, 
while the static orbit at the throat becomes unstable. 
For $\Lambda <  1.6$ only the unstable static orbit at the throat remains.

For the sequences of rotating wormhole solutions with $\omega_s>0.2$, we considered in our study, 
no stable static orbits were found.

\section{Conclusions}

In this paper we have considered wormhole solutions
supported by a complex phantom field with a Mexican
hat type potential in Einstein gravity.
The complex phantom field allows for an explicit 
dependence on time and azimuthal angle,
while still retaining a stationary axially symmetric metric.
The $U(1)$ invariance of the model gives rise to
a conserved current and an associated conserved
charge, the particle number.

Analogous to boson stars and wormholes immersed
in ordinary bosonic matter,
the harmonic time-dependence includes a boson
frequency $\omega_s$, whereas the angular-dependence
involves a winding number $n$,
and the angular momentum turns out to be proportional to 
the particle number with proportionality constant $n$.

However, the analogy between the known systems based
on ordinary complex boson fields and those based on
a complex phantom field considered here
does not carry much further.
In particular, in the presence of the complex phantom field
there arise static solutions,
i.e., solutions where the boson frequency vanishes,
which are non-singular everywhere and asymptotically flat
and possess a finite mass.

In the case of rotation, where $n$ assumes a finite value,
these solutions with vanishing boson frequency give rise
to deformed static wormholes. Here the $n$-dependence
results in an explicit dependence of the phantom field
and the metric on the polar angle.
For finite boson frequency the rotation of the phantom
field drags the throat and the spacetime along,
allowing for symmetric rotating wormholes.

For $n=0$ the wormhole solutions exist for a large range
of values of the self-interaction of the phantom field.
Indeed, the coupling constant can be made arbitrarily 
large. In contrast, the $n>0$ wormholes solutions seem
to exist only in a small interval of the coupling constant,
which decreases with increasing boson frequency.

As the coupling constant increases for a fixed boson frequency,
at a certain point the solutions change rapidly. Their mass
decreases steeply, their particle number and angular momentum
increase steeply together with the circumferential radius of the throat.
At the same time the angular frequency in the equatorial plane
tends towards zero. Unfortunately, the accuracy of the solutions
then deteriorates, and we cannot draw a reliable conclusion
on the limiting behavior.

Concerning the stability of these wormhole spacetimes
we recall that in the non-rotating case they have been shown
to possess an unstable radial mode
\cite{Dzhunushaliev:2017syc},
in complete analogy to the static Ellis wormholes
\cite{Shinkai:2002gv,Gonzalez:2008wd,Gonzalez:2008xk,Torii:2013xba}.
While for rotating wormholes in four spacetime dimensions,
such an analysis would be much more involved,
one might consider to study the rotating case first in five dimensions,
where it has been shown,
that the notorius radial instability may disappear 
\cite{Dzhunushaliev:2013jja}.


\section*{Acknowledgments}

We would like to acknowledge support by the DFG Research Training Group 1620
{\sl Models of Gravity} as well as by 
the COST Action CA16104 {\sl GWverse}.
We are grateful to the Research Group Linkage Programme of the 
Alexander von Humboldt Foundation for the support of this research.
XYC would like to thank Lucas G. Collodel and Jose Luis Bl\'{a}zquez-Salcedo for useful discussion.

\end{document}